\documentclass[prb,onecolumn,showpacs,preprintnumbers,amsmath,amssymb,floatfix]{revtex4}

\usepackage{graphicx}

\begin{document}

\title{Finite-size errors in continuum quantum Monte Carlo calculations}

\author{N.\ D.\ Drummond}

\author{R.\ J.\ Needs}

\affiliation{TCM Group, Cavendish Laboratory, University of Cambridge, J.\ J.\
Thomson Avenue, Cambridge CB3 0HE, United Kingdom}

\author{A.\ Sorouri}  \altaffiliation[Current address: ]{Department of
Physics, University of Kurdistan, P.O.\ Box 416, Pasdaran Blvd., Sanandaj
66135, Iran.}

\author{W.\ M.\ C.\ Foulkes}

\affiliation{Department of Physics, Imperial College London,  London SW7 2AZ,
United Kingdom}

\date{\today}

\begin{abstract}

We analyze the problem of eliminating finite-size errors from quantum Monte
Carlo (QMC) energy data.  We demonstrate that both (i) adding a recently
proposed\cite{fin_chiesa} finite-size correction to the Ewald energy and (ii)
using the model periodic Coulomb (MPC)
interaction\cite{fin_louisa,fin_paul,fin_rc} are good solutions to the problem
of removing finite-size effects from the interaction energy in cubic systems,
provided the exchange-correlation (XC) hole has converged with respect to
system size.  However, we find that the MPC interaction distorts the XC hole
in finite systems, implying that the Ewald interaction should be used to
generate the configuration distribution.  The finite-size correction of Ref.\
\onlinecite{fin_chiesa} is shown to be incomplete in systems of low symmetry.
Beyond-leading-order corrections to the kinetic energy are found to be
necessary at intermediate and high densities, and we investigate the effect of
adding such corrections to QMC data for the homogeneous electron gas. We
analyze finite-size errors in two-dimensional systems and show that the
leading-order behavior differs from that which has hitherto been supposed.  We
compare the efficiency of different twist-averaging methods for reducing
single-particle finite-size errors and we examine the performance of various
finite-size extrapolation formulas. Finally, we investigate the system-size
scaling of biases in diffusion QMC\@.

\end{abstract}


\maketitle

\section{Introduction}

Continuum quantum Monte Carlo (QMC) techniques\cite{foulkes_2001} enable the
total energies of many-electron systems to be calculated to very high
accuracy.  QMC simulations of condensed matter are usually performed using
finite simulation cells subject to periodic boundary conditions.  The energy
per particle is calculated at several different system sizes and the results
are extrapolated to infinite system size.  Unfortunately, this process
introduces errors into the QMC results.  Indeed, for simple systems such as
the homogeneous electron gas (HEG), finite-size extrapolation is believed to
be the largest single source of error in QMC data.

In this paper we address a number of outstanding problems associated with
finite-size extrapolation. We discuss the physics of finite-size effects in
Sec.\ \ref{sec:physics}.  In Sec.\ \ref{sec:twistav_comp} we discuss the use
of twist-averaged boundary conditions\cite{lin_twist_2001} to reduce errors
caused by momentum quantization in finite simulation cells.  In Sec.\
\ref{sec:mpc_equiv_fscorr} we give results illustrating that recently proposed
methods for correcting the Ewald energy\cite{fin_chiesa,fin_gaudoin} are
essentially equivalent to the use of the model periodic Coulomb (MPC)
interaction\cite{fin_louisa,fin_paul,fin_rc} in QMC simulations.  In Sec.\
\ref{sec:nonanalytic} we discuss various complications posed by low-symmetry
systems.  In Sec.\ \ref{sec:get_ke_right} we demonstrate that the finite-size
correction to the kinetic energy (KE) proposed in Ref.\
\onlinecite{fin_chiesa} is incomplete and that higher-order terms cannot be
neglected at typical metallic densities.  We analyze finite-size errors in
2D-periodic systems in Sec.\ \ref{sec:fs_2D}.  In Sec.\
\ref{sec:extrapolation_expressions} we investigate the performance of
different finite-size extrapolation formulas. In Sec.\ \ref{sec:biases} we
examine the size-dependence of biases in QMC energies.  Finally we draw our
conclusions in Sec.\ \ref{sec:conclusions}.

Hartree atomic units (a.u.)\ are used throughout, in which the Dirac constant,
the magnitude of the electronic charge, the electronic mass, and $4\pi$ times
the permittivity of free space are unity: $\hbar = |e| = m_e = 4 \pi
\epsilon_0 = 1$.  All our QMC calculations were carried out using the
\textsc{casino} code.\cite{casino} We have made use of the variational and
diffusion quantum Monte Carlo (VMC and DMC) methods.\cite{foulkes_2001}
Throughout, we specify the density of a HEG by quoting the radius $r_s$ of the
sphere (circle in 2D) that contains one electron on average.

\section{Physics of finite-size effects \label{sec:physics}}

\subsection{Components of the total energy}

The total energy of a periodic many-electron system can be divided into (i)
the KE, (ii) the electron-electron interaction energy, and (iii) the
electron-ion interaction energy (we include the interaction of the electrons
with any other external fields in this term).  The electron-electron
interaction energy may be subdivided into the Hartree and exchange-correlation
(XC) energies.  Assuming the electrostatic potential to be periodic, the
former is the Coulomb energy due to the periodic charge density, and the
latter is the remainder of the electron-electron interaction energy, which
arises from the correlation of electron motions and the antisymmetry of the
many-electron wave function.  The electron-ion interaction energy and the
Hartree energy depend only on the electronic charge density, which has the
periodicity of the primitive cell and is rapidly convergent with respect to
system size; hence the finite-size errors in these energy components are
generally negligible.  By contrast, the finite-size errors in the XC energy
and the KE can be very substantial. We analyze the physics underlying these
errors in the rest of this section.

\subsection{Simulation and primitive unit cells for crystalline solids}

Suppose we wish to calculate the energy per particle of a periodic solid.  In
one-electron theories we can often reduce the problem to the primitive unit
cell and integrate over the first Brillouin zone.  Reduction to the primitive
cell is not possible in many-body calculations because correlation effects may
be long-ranged, and hence such calculations must be performed in simulation
cells consisting of several primitive cells. Periodic boundary conditions are
imposed across the simulation cell.

Suppose the simulation cell contains $N$ electrons and let $\{ {\bf
r}_1,\ldots,{\bf r}_N\}$ be the electron coordinates.  The simulation-cell
Hamiltonian $\hat{H}$ satisfies
\begin{eqnarray}
\hat{H}({\bf r}_1,\ldots,{\bf r}_i+{\bf R}_s,\ldots,{\bf r}_N) & = &
  \hat{H}({\bf r}_1,\ldots,{\bf r}_i,\ldots,{\bf r}_N) ~~~~~ \forall i \in
  \{1,\ldots,N\} \\ \hat{H}({\bf r}_1+{\bf R}_p,\ldots,{\bf r}_i+{\bf
  R}_p,\ldots,{\bf r}_N+{\bf R}_p) & = & \hat{H}({\bf r}_1,\ldots,{\bf
  r}_i,\ldots,{\bf r}_N),
\end{eqnarray}
where ${\bf R}_s$ and ${\bf R}_p$ are simulation-cell and primitive-cell
lattice vectors.  The first of these symmetries is an artifact of the
periodicity imposed on the simulation cell.  These translational symmetries
lead to the many-body Bloch conditions
\begin{eqnarray}
\Psi_{{\bf k}_s}({\bf r}_1,\ldots,{\bf r}_N) & = & U_{{\bf k}_s}({\bf
  r}_1,\ldots,{\bf r}_N) \exp \left( i {\bf k}_s \cdot \sum_i {\bf r}_i
  \right) \\ \Psi_{{\bf k}_p}({\bf r}_1,\ldots,{\bf r}_N) & = & W_{{\bf
  k}_p}({\bf r}_1,\ldots,{\bf r}_N) \exp \left( i {\bf k}_p \cdot \frac{1}{N}
  \sum_i {\bf r}_i \right),
\end{eqnarray}
where $U$ has the periodicity of the simulation cell for every electron and
$W$ is invariant under the simultaneous translation of all electrons through
${\bf R}_p$.\cite{rajagopal_1994,rajagopal_1995} The use of a nonzero
simulation-cell Bloch vector ${\bf k}_s$ is sometimes described as the
application of \textit{twisted} boundary conditions.\cite{lin_twist_2001} The
center-of-mass Bloch momentum ${\bf k}_p$ may be restricted to the Brillouin
zone corresponding to the primitive lattice, while the twist vector ${\bf
k}_s$ may be restricted to the smaller Brillouin zone corresponding to the
simulation-cell lattice.  From now on, we use ${\bf G}_s$ and ${\bf G}_p$ to
denote vectors in the simulation-cell and primitive-cell reciprocal lattices,
respectively.

\subsection{Single-particle finite-size errors}

In a finite simulation cell subject to periodic boundary conditions, each
single-particle orbital can be taken to be of Bloch form $\psi_{\bf k}({\bf
r})=\exp[i{\bf k}\cdot{\bf r}] u_{\bf k}({\bf r})$, where $u_{\bf k}$ has the
periodicity of the primitive cell and ${\bf k}$ lies on the grid of integer
multiples of the simulation-cell reciprocal-lattice vectors within the first
Brillouin zone of the primitive cell, the grid being offset from the origin by
${\bf k}_s$, so that ${\bf k} = {\bf k}_s + {\bf G}_s$ for some ${\bf G}_s$.
Instead of integrating over single-particle orbitals inside the Fermi surface
to calculate the Hartree--Fock (HF) KE and exchange energy, one therefore sums over a discrete
set of ${\bf k}$ vectors when a finite cell is used.  For metallic systems,
the set of occupied ground-state orbitals depends on ${\bf k}_s$ and hence
calculated properties are nonanalytic functions of ${\bf k}_s$.  As the system
size is increased, the fineness of the grid of single-particle Bloch ${\bf k}$
vectors increases and the HF energy changes abruptly as shells of orbitals
pass through the Fermi surface.

Fluctuations in the QMC KE contain large ``single-particle'' contributions
that are roughly proportional to the corresponding fluctuations in the HF
KE\@. Hence HF energy data can be used to extrapolate QMC energies to infinite
system size, as discussed in Sec.\ \ref{sec:extrapolation_expressions}.  Note
that a judicious choice of ${\bf k}_s$ (e.g., the Baldereschi
point\cite{baldereschi} for insulators) can greatly reduce single-particle
finite-size errors.\cite{rajagopal_1994,rajagopal_1995} The common choice of
${\bf k}_s={\bf 0}$ generally maximizes shell-filling effects and is usually
the worst possible value for estimating the total energy, although it does
ensure that the wave function of the finite simulation cell can be chosen to
have the full symmetry of the Hamiltonian.

\subsection{Twist averaging \label{subsec:twist-averaging}}

\textit{Twist averaging} within the canonical ensemble (CE) means taking the
average of expectation values over all simulation-cell Bloch vectors ${\bf
k}_s$ in the first Brillouin zone of the simulation cell, i.e., over all
offsets to the grid of ${\bf k}$ vectors, keeping the number of electrons
fixed.\cite{lin_twist_2001}

At the HF level, the effect of twist averaging within the CE is to replace
sums over the discrete set of single-particle orbitals by integrals over a
volume of ${\bf k}$-space.  Consider, for example, a simulation cell of HEG
containing an even number of electrons $N$.  For each twist ${\bf k}_s$, the
$N/2$ shortest Bloch vectors of the form ${\bf k}_s + {\bf G}_s$ are doubly
occupied. Integrating over twists therefore averages over the volume of ${\bf
k}$-space occupied by the first $N/2$ Brillouin zones of the simulation
cell. The occupied region is a convex polyhedron that tends to the Fermi
surface in the limit of infinite system size and has the correct volume at all
system sizes.  Since the single-particle KE $k^2/2$ is a convex function of
${\bf k}$, the small differences between the occupied region of ${\bf
k}$-space and the Fermi volume cause the CE twist-averaged HF KE to be
slightly too large for finite systems. This systematic error, which exhibits
visible shell-filling effects, decays with system size.

Twist averaging within the grand canonical ensemble (GCE) also means taking
the average of expectation values with respect to ${\bf k}_s$, but this time
allowing the number of electrons to vary with ${\bf k}_s$.  For any given
${\bf k}_s$, only those states that lie within the Fermi surface are occupied.
This allows one to integrate over the Fermi volume in simulations with a
finite number of particles, so that the HF KE of a HEG is exact at all system
sizes.  The KE at a given ${\bf k}_s$ is obtained by summing the one-electron
KE's of the occupied states. Values of ${\bf k}_s$ with fewer occupied states
therefore contribute less to the GCE average.

We compare the efficiency of grid-based and Monte Carlo methods for
integrating over the simulation-cell Bloch vector ${\bf k}_s$ in Sec.\
\ref{sec:twistav_comp}, where we also discuss the use of the CE and GCE in HF
calculations.

\subsection{Ewald interaction}

When simulating infinite periodic systems or finite systems subject to
periodic boundary conditions, it is not possible to use the familiar $1/r$
form of the Coulomb interaction because the sums over images of the simulation
cell do not converge absolutely. The standard solution to this problem is to
replace the Coulomb interaction by the Ewald interaction.\cite{ewald} The 3D
Ewald interaction is the periodic solution of Poisson's equation for a
periodic array of point charges embedded in a uniform neutralizing background
and is therefore appropriate for an electrically unpolarized, neutral
system. Using the 3D Ewald interaction corresponds to adding a neutralizing
background if necessary and calculating the Coulomb energy per simulation cell
of a macroscopic array of identical copies of the simulation cell embedded in
a perfect metal so that surface polarization charges are always
screened.\cite{fin_louisa} The Ewald energy for any particular electron
configuration in a 3D system is
\begin{equation}
\label{eq:VEw}
\hat{V}_{\rm Ew} = \frac{1}{2} \sum_{i\neq j} v_E({\bf r}_i-{\bf r}_j) +
\frac{1}{2} N v_M,
\end{equation}
where $v_E({\bf r})$ is the Ewald interaction and $v_M \equiv \lim_{{\bf r}
\rightarrow {\bf 0}} \left[ v_E({\bf r}) - 1/r \right]$ is the Madelung
constant, which represents the interaction between a point charge and its own
images and canceling background. These quantities may be evaluated efficiently
using the Ewald formulas:
\begin{eqnarray}
v_E({\bf r}) & = & \frac{1}{\Omega} \sum_{{\bf G}_s \neq {\bf 0}}
\frac{4\pi\exp\left(-\kappa^2 G_s^2/2 + i{\bf G}_s\cdot{\bf r} \right
)}{G_s^2} - \frac{2\pi\kappa^2}{\Omega} + \sum_{{\bf R}_s} \frac{\mbox{erfc}
\left[ |{\bf r} - {\bf R}_s|/(\sqrt{2}\kappa) \right]} {|{\bf r} - {\bf R}_s|}
,
\label{eq:vedef} \\
v_M & = & \frac{1}{\Omega}\sum_{{\bf G}_s \neq {\bf 0}}
\frac{4\pi\exp[-G_s^2/(4\kappa^2)]}{G_s^2} - \frac{\pi}{\kappa^2 \Omega} +
\sum_{{\bf R}_s \neq {\bf 0}} \frac{{\rm erfc}(\kappa R_s)}{R_s}  -
\frac{2\kappa}{\sqrt{\pi}} ,
\label{eq:vmdef}
\end{eqnarray}
where $\Omega$ is the volume of the simulation cell.  The value of the
constant $\kappa$ does not affect $v_E({\bf r})$ or $v_M$ and may be chosen to
maximize computational efficiency. The zero of potential has been chosen such
that $v_E({\bf r})$ averages to zero over the simulation cell.  The periodic
function $v_E({\bf r})$ has Fourier
components\cite{fin_louisa,footnote:ft_conv} $v_E({\bf G}_s) = 4\pi/G_s^2$ for
${\bf G}_s \neq {\bf 0}$ and $v_E({\bf G}_s) = 0$ for ${\bf G}_s = {\bf 0}$.
Setting $\kappa = 1/(2\sqrt{\epsilon})$, where $\epsilon$ is very small, Eq.\
(\ref{eq:vmdef}) gives
\begin{equation}
\label{eq:vmapprox}
v_M \approx \frac{1}{\Omega} \sum_{{\bf G}_s \neq {\bf 0}} \frac{4\pi \exp( -
\epsilon G_s^2 ) }{G_s^2} - \frac{1}{\sqrt{\pi\epsilon}} = \frac{1}{\Omega}
\sum_{{\bf G}_s \neq {\bf 0}} \frac{4\pi \exp( - \epsilon G_s^2 ) }{G_s^2} -
\frac{1}{(2\pi)^3} \int_{k<\infty}  \frac{4\pi \exp( - \epsilon k^2 ) }{k^2}
\, d{\bf k},
\end{equation}
which will prove useful later on.

The analogous expression for the quasi-2D Ewald interaction is obtained by
solving the 3D Poisson's equation for a 2D-periodic lattice of point charges
subject to periodic boundary conditions in the plane and symmetric boundary
conditions perpendicular to the plane, and is thus appropriate for planar and
slab systems.\cite{wood} When evaluated in the plane of the charges,
$r_{\perp} = 0$, the 2D Fourier components $v_E({\bf
G}_{s\parallel},r_{\perp})$ of the quasi-2D Ewald interaction $v_E({\bf
r}_{\parallel},r_{\perp})$ are equal to $2\pi/G_{s\parallel}$ for ${\bf
G}_{s\parallel} \neq {\bf 0}$ and to 0 for ${\bf G}_{s\parallel} = {\bf 0}$.

\subsection{Structure factor and XC hole}

The analysis of the Coulomb and KE finite-size effects is most easily
expressed in terms of the static structure factor (SF), the pair density, and
the XC hole. The definitions of these quantities and relations between them
are reviewed in this section.

The SF is
\begin{equation}
S({\bf r},{\bf r}^\prime) = \frac{\Omega}{N} \left< [\hat{\rho}({\bf
r})-\rho({\bf r}) ][ \hat{\rho}({\bf r}^\prime)-\rho({\bf r}^\prime)] \right>
= \frac{\Omega}{N}\left [ \left < \hat{\rho}({\bf r})\hat{\rho}({\bf
r}^\prime) \right > - \rho({\bf r}) \rho({\bf r}^\prime) \right ],
\end{equation}
where $\hat{\rho}({\bf r})=\sum_i \delta({\bf r}-{\bf r}_i)$ is the operator
for the electron number density at position ${\bf r}$, and $\rho({\bf r})
=\langle \hat{\rho}({\bf r}) \rangle$ is its expectation value.  In periodic
systems, the Dirac delta functions are to be interpreted periodically:
$\delta[{\bf r} - ({\bf r}_i + {\bf R}_s)] = \delta({\bf r} - {\bf r}_i)$.
The SF is closely related to the pair density defined by
\begin{equation}
\rho_2({\bf r},{\bf r}^\prime) =   \left< \sum_{i \neq j} \delta({\bf r}-{\bf
r}_i)  \delta({\bf r}^\prime-{\bf r}_j) \right>  = \frac{N}{\Omega} S({\bf
r},{\bf r}^\prime) + \rho({\bf r})\rho({\bf r}^\prime) - \delta({\bf r}-{\bf
r}^\prime)\rho({\bf r}^\prime). \label{eq:rho2def}
\end{equation}
Another related quantity is the XC hole, $\rho_{\rm xc}({\bf r},{\bf
  r}^\prime)$, defined by
\begin{equation}
\rho_{\rm xc}({\bf r},{\bf r}^\prime)\rho({\bf r}^\prime) = \rho_2({\bf
r},{\bf r}^\prime) - \rho({\bf r})\rho({\bf r}^\prime) =
\frac{N}{\Omega}S({\bf r},{\bf r}^\prime) - \delta({\bf r}-{\bf
r}^\prime)\rho({\bf r}^\prime).
\end{equation}
Integrating Eq.\ (\ref{eq:rho2def}) with respect to ${\bf r}$ yields
$\int_{\Omega} \rho_2({\bf r},{\bf r}') \, d{\bf r} = (N-1) \langle \sum_j
\delta({\bf r}' - {\bf r}_j)\rangle = (N-1)\rho({\bf r}')$ and hence we obtain
the sum rule $\int_{\Omega} \rho_{\rm xc}({\bf r},{\bf r}^\prime) \, d{\bf r}
= -1$.  The XC hole describes the suppression of the electron density at ${\bf
r}$ caused by the presence of an electron at ${\bf r}^\prime$.

It is often more convenient to work with the translationally averaged SF
\begin{equation}
S({\bf r})=\frac{1}{\Omega} \int_\Omega S({\bf r}^\prime+{\bf r},{\bf
r}^\prime)\, d{\bf r}^\prime ,
\end{equation}
and the analogous translationally averaged pair density $\rho_2({\bf r})$.
  These quantities have the periodicity of the simulation cell and may be
  expanded as Fourier series, the components of which are
\begin{eqnarray}
S({\bf G}_s) & = & \frac{1}{N} \left[ \left< \hat{\rho}({\bf G}_s)
\hat{\rho}^\ast({\bf G}_s) \right> - \rho({\bf G}_s) \rho^\ast({\bf G}_s)
\right], \\ \rho_2({\bf G}_s) & = & \frac{1}{\Omega}\left < \hat{\rho}({\bf
G}_s) \hat{\rho}^{\ast}({\bf G}_s)\right > - \frac{N}{\Omega}   =
\frac{N}{\Omega}\left [ S({\bf G}_s) - 1 \right ] + \frac{1}{\Omega} \rho({\bf
G}_s)\rho^{\ast}({\bf G}_s) ,
\end{eqnarray}
where $\hat{\rho}({\bf G}_s)=\sum_i \exp(-i{\bf G}_s \cdot {\bf r}_i)$ is a
  Fourier component of the density operator.\cite{footnote:ft_conv} Finally,
  the system-averaged XC hole is defined as
\begin{equation}
\rho_{\rm xc}({\bf r}) = \frac{1}{N}\int_\Omega \rho_{\rm xc}({\bf r}^\prime +
  {\bf r}, {\bf r}^\prime) \rho({\bf r}^\prime) \, d{\bf r}^\prime = S({\bf
  r}) - \delta({\bf r}). \label{eq:rhoxcSrel}
\end{equation}

\subsection{Hartree and XC energies}

The Ewald interaction energy is the expectation value of the operator in Eq.\
(\ref{eq:VEw}):
\begin{eqnarray}
\left< \hat{V}_{\rm Ew} \right>  & = &  \frac{N v_M}{2} + \frac{1}{2}
\int_\Omega \int_\Omega \left < \sum_{i \neq j}  \delta({\bf r} - {\bf
r}_i)\delta({\bf r}^\prime-{\bf r}_j)  \right > v_E({\bf r} - {\bf r}^\prime)
\, d{\bf r} \, d{\bf r}^\prime \nonumber \\ & = & \frac{N v_M}{2} +
\frac{1}{2} \int_{\Omega} \int_{\Omega} \rho_2({\bf r},{\bf r}^\prime) \,
v_E({\bf r} - {\bf r}^\prime) \, d{\bf r} \, d{\bf r}^\prime
\label{eqn:rs_rho2} \\ & = & \frac{1}{2} \int_\Omega \int_\Omega
\rho_{\rm xc}({\bf r},{\bf r}^\prime) \rho({\bf r}^\prime) \left[ v_E({\bf
r}-{\bf r}^\prime) - v_M \right]\, d{\bf r} \, d{\bf r}^\prime +
\frac{1}{2}\int_\Omega \int_\Omega \rho({\bf r}) \rho({\bf r}^\prime) v_E({\bf
r}-{\bf r}^\prime) \, d{\bf r} \, d{\bf r}^\prime
\label{eqn:rs_har_p_xc}  \\ & = &
\frac{N}{2} \left( v_M + \frac{1}{\Omega} \sum_{{\bf G}_s \neq {\bf 0}}
v_E(G_s) \left[ S({\bf G}_s)-1 \right] \right) + \frac{1}{2\Omega} \sum_{{\bf
G}_p \neq {\bf 0}} v_E(G_p) \rho({\bf G}_p) \rho^\ast({\bf G}_p),
\label{eqn:ks_har_p_xc}
\end{eqnarray}
where use has been made of the sum rules $\int_\Omega \rho_{\rm xc}({\bf
r},{\bf r}^\prime) \, d{\bf r} = -1$ and $\int_\Omega \rho({\bf r}^\prime) \,
d{\bf r}^\prime = N$.  The first term in Eqs.\ (\ref{eqn:rs_har_p_xc}) and
(\ref{eqn:ks_har_p_xc}) is the XC energy (the interaction of the electrons
with their XC holes).\cite{footnote:xc_energy} The second term is the Hartree
energy (the interaction of the charge densities).  The Hartree term in Eq.\
(\ref{eqn:ks_har_p_xc}) has been simplified by noting that $\rho({\bf r})$ has
the periodicity of the primitive lattice and hence that $\rho({\bf G}_s)$
vanishes unless ${\bf G}_s \in \{ {\bf G}_p \}$.

In practice the charge density and pair density converge rapidly with system
size for interacting systems,\cite{fin_rc} due to the fact that the XC hole
falls off very quickly with $r$.  For example, the nonoscillatory part of the
XC hole falls off as $r^{-8}$ for a 3D HEG\@.\cite{gorigiorgi_perdew} If the
charge density is correct then the Hartree energy in a finite cell is exact,
as can be seen from Eq.\ (\ref{eqn:ks_har_p_xc}): the Fourier components
$v_E(G_p)$ are equal to $4\pi/G_p^2$ and $\rho({\bf G}_p)$ is proportional to
the number of primitive cells in $\Omega$, so the Hartree energy per electron
is independent of system size. The finite-size errors in the interaction
energy given by Eq.\ (\ref{eqn:rs_har_p_xc}) must therefore be caused by the
slow convergence of the Ewald interaction $v_E({\bf r})-v_M$ in the XC energy.

A power expansion of the Ewald interaction about ${\bf r} = {\bf 0}$
gives\cite{fin_louisa}
\begin{equation}
v_E({\bf r}) - v_M = \frac{1}{r} + \frac{2 \pi}{3 \Omega} {\bf r}^{\rm T} W
{\bf r} +O\left( \frac{r^4}{\Omega^{5/3}} \right),
\label{eqn:taylor_ewald}
\end{equation}
where the tensor $W$ depends on the symmetry of the lattice. $W$ is the
identity matrix for a lattice of cubic symmetry.  For large simulation cells
the first term in the expansion dominates in the region where the XC hole is
large, but for typical cell sizes the second term can be significant.  Unlike
the Hartree energy, we do not want the effect of periodic images in the XC
energy: the interaction between each electron and its XC hole should just be
$1/r$.  This is enforced in the MPC
interaction.\cite{fin_louisa,fin_paul,fin_rc}

In HF theory, unlike QMC and reality, the exchange hole is long-ranged (the
nonoscillatory tail falls off as $r^{-4}$) and the pair density is slowly
convergent with system size.\cite{giuliani} This gives an additional source of
finite-size error, even when the MPC interaction is used, as discussed in
Appendix \ref{sec:hf_fs_errors}.

\subsection{MPC interaction \label{sec:mpc}}

The MPC interaction operator\cite{fin_louisa,fin_paul,fin_rc} is
\begin{eqnarray}
\hat{V}_{\rm MPC} & = & \frac{1}{2} \sum_{i\neq j} f({\bf r}_i-{\bf r}_j) +
\sum_i \int_\Omega \rho({\bf r}) \left[ v_E({\bf r}_i-{\bf r})- f({\bf
r}_i-{\bf r})\right] \, d{\bf r} \nonumber \\ && - \frac{1}{2} \int_\Omega
\rho({\bf r})\rho({\bf r}^{\prime}) \left[v_E({\bf r}-{\bf r}^{\prime}) -
f({\bf r}-{\bf r}^{\prime})\right] \, d{\bf r} \, d{\bf r}^{\prime},
\end{eqnarray}
where $f({\bf r})$ is $1/r$ treated within the minimum-image convention in the
simulation cell.\cite{allen} Assuming the pair density  and the charge density
have converged to their infinite-system  forms, the MPC electron-electron
interaction energy is
\begin{equation}
\left< \hat{V}_{\rm MPC} \right> = \frac{1}{2} \int_\Omega \int_\Omega
\rho_{\rm xc}({\bf r},{\bf r}^\prime) \rho({\bf r}^{\prime}) f({\bf r}-{\bf
r}^\prime) \, d{\bf r}\,d{\bf r}^\prime  + \frac{1}{2}\int_\Omega \int_\Omega
\rho({\bf r})\rho({\bf r}^{\prime}) v_E({\bf r}-{\bf r}^{\prime}) \, d{\bf r}
\, d{\bf r}^{\prime}. \label{eqn:MPC_int_energy}
\end{equation}
Hence the Hartree energy is calculated using the Ewald interaction while the
XC energy is calculated using $1/r$ (within the minimum-image convention), as
desired.  The MPC interaction energy per electron therefore converges more
rapidly with system size than the Ewald interaction energy.  One can avoid the
need to know $\rho$ exactly by replacing it with the approximate (but usually
highly accurate) charge density $\rho_{\rm A}$ from a
density-functional-theory (DFT) or HF calculation in $\hat{V}_{\rm MPC}$.  The
error due to this approximation is $O(\rho-\rho_{\rm A})^2$.  Comparing Eqs.\
(\ref{eqn:MPC_int_energy}) and (\ref{eqn:rs_har_p_xc}), we see that the
difference between the Ewald and MPC XC energies involves the operator $(v_E -
v_M - f)$, which vanishes as the size of the simulation cell goes to
infinity. So the Ewald and MPC XC energies per particle are the same in the
limit of large system size, even if an approximate charge density is used.  In
practice the first term of the MPC interaction is evaluated in real space, the
second term is evaluated in ${\bf k}$-space, and the third term is a
constant:\cite{maezono}
\begin{eqnarray}
\hat{V}_{\rm MPC} & = & \frac{1}{2} \sum_{i\neq j} f({\bf r}_i-{\bf r}_j) +
\frac{1}{\Omega} \sum_i \sum_{{\bf G}_p\neq {\bf 0}} \left[ v_E(G_p) - f({\bf
G}_p) \right] \rho_{\rm A}({\bf G}_p) \exp(i{\bf G}_p\cdot {\bf r}_i)
\nonumber \\ & & {} + \left(- \frac{1}{\Omega} N f_{{\bf 0}} \rho_{{\rm A}{\bf
0}} - \frac{1}{2\Omega} \sum_{{\bf G}_p\neq {\bf 0}} \left[ v_E(G_p) - f({\bf
G}_p) \right] \rho^*_{\rm A}({\bf G}_p) \rho_{\rm A}({\bf G}_p) +
\frac{1}{2\Omega} f_{\bf 0} \rho^*_{{\rm A}{\bf 0}} \rho_{{\rm A}{\bf 0}}
\right),
\end{eqnarray}
where $f_{\bf 0}$ and $\rho_{{\rm A}{\bf 0}}$ are the ${\bf G}_s={\bf 0}$
components of $f$ and $\rho_{\rm A}$.  Although $f({\bf r})$ has the
periodicity of the simulation cell, its Fourier components are only required
on primitive lattice vectors ${\bf G}_p$. These Fourier components are
evaluated numerically in advance, a procedure that requires some care because
$f({\bf r})$ diverges at ${\bf r}={\bf 0}$ and is nondifferentiable at the
boundary of the Wigner-Seitz cell of the simulation cell.  Once the Fourier
components have been obtained, the MPC interaction is much quicker to evaluate
than the Ewald interaction because (i) there is no real-space sum over lattice
vectors and (ii) the ${\bf k}$-space sum runs over primitive-cell ${\bf G}_p$
vectors only, so the number of ${\bf G}_p$ vectors to include in the sum does
not grow with system size.

\subsection{Finite-size correction to the Ewald energy in 3D
\label{subsec:fscee}}

Assuming that the charge density (and hence Hartree energy) and the Fourier
components of the SF converge rapidly with system size, it follows by
comparing Eq.\ (\ref{eqn:ks_har_p_xc}) with its infinite system-size limit
that the finite-size correction to the 3D Ewald interaction energy is
\begin{equation}
\Delta V_{\rm Ew} = \frac{N}{2} \left( \frac{1}{(2 \pi)^3} \int_{k<\infty}
v_E(k) \left[ {S}({\bf k})-1 \right] \, d{\bf k} - \frac{1}{\Omega} \sum_{{\bf
G}_s \neq {\bf 0}} v_E(G_s) \left[ {S}({\bf G}_s)-1 \right] - v_M \right) ,
\end{equation}
where we have noted that $v_M \rightarrow 0$ as the system size tends to
infinity.  Since ${S}({\bf k}) \rightarrow 1$ as $k \rightarrow \infty$, the
sum and the integral converge, allowing us to include factors of
$\exp(-\epsilon k^2)$ in the summand and integrand without affecting $\Delta
V_{\rm Ew}$ if $\epsilon$ is small enough. Substituting for $v_M$ using Eq.\
(\ref{eq:vmapprox}) then gives
\begin{equation}
\Delta V_{\rm Ew} \approx \frac{N}{2} \left( \frac{1}{(2 \pi)^3}
\int_{k<\infty} v_E(k) {S}({\bf k}) \exp \left(-\epsilon k^2 \right) \, d{\bf
k}  - \frac{1}{\Omega} \sum_{{\bf G}_s \neq {\bf 0}} v_E(G_s) {S}({\bf G}_s)
\exp \left( -\epsilon G_s^2 \right) \right). \label{eq:dvew2}
\end{equation}
The convergence factors are now required to keep the summation and integration
finite, even though they do not affect the value of the expression as a whole.

An obvious contribution to the finite-size correction is apparent from the
form of Eq.\ (\ref{eq:dvew2}).  In interacting electron systems with cubic (or
higher) symmetry, ${S}({\bf k})= \eta k^2 + O(k^4)$ for small $k$, where
$\eta$ is a constant.\cite{wang} The function $v_E(k) S({\bf k}) = 4 \pi
S({\bf k}) / k^2$ therefore tends to a well-defined limit as $k\rightarrow 0$,
suggesting that much of the difference between the sum and the integral in
Eq.\ (\ref{eq:dvew2}) is caused by the omission of the ${\bf G}_s = {\bf 0}$
term from the summation.  This argument leads to a finite-size correction of
the form derived by Chiesa \textit{et al.}:\cite{fin_chiesa}
\begin{equation}
\Delta V_{\rm Ew} \approx \frac{N}{2\Omega} \lim_{k\rightarrow 0} \frac{4\pi
S({\bf k})}{k^2} = \frac{2\pi N \eta}{\Omega}.
\label{eq:finsize1}
\end{equation}
Since $\langle \hat{V}_{\rm Ew} \rangle$ is proportional to system size, the
relative error in the Ewald energy falls off as $O(N^{-1})$.  In a 3D HEG, the
random phase approximation (RPA) implies that $\eta=1/(2 \omega_p)$, where
$\omega_p=\sqrt{4 \pi N/\Omega} = \sqrt{3/r_s^3}$ is the plasma
frequency,\cite{wang,pines} giving\cite{fin_chiesa}
\begin{equation}
\Delta V_{\rm Ew}=\frac{\omega_p}{4}.
\label{eqn:xc_corr_heg}
\end{equation}

These approximate arguments may be made more precise and given an appealing
physical interpretation as follows. According to Eq.\ (\ref{eq:rhoxcSrel}),
$S({\bf r}) = \rho_{\rm xc}({\bf r}) + \delta({\bf r})$ can be viewed as the
localized charge distribution of an electron at the origin and the
system-averaged XC hole surrounding it. More precisely, because the simulation
cell is repeated periodically, $S({\bf r})$ is a superposition of many such
localized charge distributions, one centered in every copy of the simulation
cell, i.e., $S({\bf r}) = \sum_{{\bf R}_s} S_{\rm loc}({\bf r} - {\bf R}_s)$.
If we assume that the XC hole is well localized within the simulation cell,
which must be the case if $S({\bf k})$ has converged with respect to system
size, this decomposition is unambiguous.  It is then easy to show that
\begin{equation}
S({\bf G}_s) = \int_{r < \infty} S_{\rm loc}({\bf r}) \exp(-i{\bf G}_s
\cdot{\bf r}) \, d{\bf r}.
\end{equation}
The discrete Fourier components of the periodic function $S({\bf r})$ are
therefore equal to the corresponding components of the continuous Fourier
transform of the localized function $S_{\rm loc}({\bf r})$.  If $S_{\rm
loc}({\bf r})$ is  convolved with a very narrow normalized Gaussian
$(4\pi\epsilon)^{-3/2} \exp(-r^2/4\epsilon)$ before the Fourier transform  is
taken, $S({\bf k})$ is multiplied by the convergence factor  $\exp(-\epsilon
k^2)$ appearing in Eq.\ (\ref{eq:dvew2}). The convolution smears out the delta
function slightly, but has no other discernible effect on the form of $S_{\rm
loc}({\bf r})$.

We can now interpret the two terms between the large parentheses in Eq.\
(\ref{eq:dvew2}). The integral is the value at the origin of the potential
\begin{equation}
\phi_{{\rm loc},\epsilon}({\bf r}) =  \int_{r^\prime < \infty} \frac{S_{{\rm
loc},\epsilon}({\bf r}^\prime)}{|{\bf r}-{\bf r}^\prime|} \, d{\bf r}^\prime
\end{equation}
corresponding to the aperiodic charge density $S_{{\rm loc},\epsilon}({\bf
r})$ obtained by convolving $S_{\rm loc}({\bf r})$ with the very narrow
Gaussian. The summation [including the missing ${\bf G}_s = {\bf 0}$ term,
which is well-defined for systems of cubic symmetry or if $S({\bf G}_s)$ is
replaced by its spherical average\cite{footnote:shell_spherical}] is the value
at the origin of the potential
\begin{equation}
\label{eqn:phiepsilon}
\phi_{\epsilon}({\bf r}) = \sum_{{\bf R}_s}  \phi_{{\rm loc},\epsilon}({\bf r}
- {\bf R}_s)
\end{equation}
of an infinite periodic lattice of copies of $S_{{\rm loc},\epsilon}({\bf
r})$.  The sum rule $\int_{r<\infty} S_{{\rm loc},\epsilon}({\bf r}) \, d{\bf
r}=0$ ensures that $S_{{\rm loc},\epsilon}({\bf r})$ has no monopole and the
system averaging of the symmetric function $S({\bf r},{\bf r}^{\prime}) =
S({\bf r}^{\prime},{\bf r})$ ensures that $S_{{\rm loc},\epsilon}({\bf r})$
has no dipole.\cite{footnote:nodipole} If the system has cubic symmetry or we
approximate $S_{{\rm loc},\epsilon}({\bf r})$ by its spherical average as
proposed in Ref.\ \onlinecite{fin_gaudoin}, the quadrupole vanishes too and
$\phi_{{\rm loc},\epsilon}({\bf r})$ decays rapidly enough to ensure that the
summation in Eq.\ (\ref{eqn:phiepsilon}) converges.  Equation (\ref{eq:dvew2})
can then be rewritten as
\begin{eqnarray}
\Delta V_{\rm Ew} & \approx  & \frac{N}{2} \left ( \phi_{\rm loc}({\bf 0}) -
\left [ \sum_{{\bf R}_s} \phi_{\rm loc}({\bf R}_s) - \frac{1}{\Omega}
\lim_{{\bf k} \rightarrow {\bf 0}} v_E(k)S({\bf k}) \right ] \right )
\nonumber \\ & = &  \frac{N}{2} \left( \frac{4 \pi}{\Omega} \lim_{{\bf
k}\rightarrow {\bf 0}} \frac{{S}({\bf k})}{k^2} - \sum_{{\bf R}_s \neq {\bf
0}} \phi_{\rm loc}({\bf R}_s) \right),
\label{eqn:general_xc_corr}
\end{eqnarray}
a result that can also be obtained using the Poisson summation
formula\cite{footnote:poisson} (which we have, in effect, derived). The first
term is the finite-size correction from Eq.\ (\ref{eq:finsize1}) and the
second term is small, as explained below.

The nonoscillatory tail of the spherical XC hole of a 3D HEG is of the form
$\rho_{\rm xc}(r)=-\Lambda r^{-8}$, where $\Lambda$ is a
constant.\cite{gorigiorgi_perdew} It arises from the $O(k^5)$ term in
$S(k)$.\cite{footnote:k3_term} The total XC charge lying further than $r$ from
the origin is therefore $-4\pi \Lambda/(5 r^5)$, so $\phi_{\rm loc}(r) = 4 \pi
\Lambda/(5r^6)$ for large $r$.  Hence
\begin{equation}
- \frac{N}{2} \sum_{{\bf R}_s \neq {\bf 0}} \phi_{\rm loc}({\bf R}_s) \approx
-\frac{N}{2\Omega} \int_{R_\Omega}^\infty \frac{4 \pi \Lambda}{5 r^6} 4 \pi
r^2 \, dr = O(N^{-1}),
\end{equation}
where $R_\Omega$ is the radius of a sphere of volume $\Omega$.  Thus, the
remaining error in the XC energy per particle not accounted for by Eq.\
(\ref{eq:finsize1}) is $O(N^{-2})$.

In inhomogeneous systems, $\rho_{\rm xc}({\bf r})$ may not be spherical,
causing $\phi_{\rm loc}({\bf r})$ to decay more slowly at large $r$. In
particular, if $S_{\rm loc}({\bf r})$ has a nonzero quadrupole moment,
$\phi_{\rm loc}({\bf r}) \propto r^{-3}$ and the sum over ${\bf R}_s$ fails to
converge absolutely.  The error not accounted for by the XC correction
proposed by Chiesa \textit{et al.}\cite{fin_chiesa}\ is then of the same order
as the correction itself.  These additional errors are related to the behavior
of $S({\bf k})/k^2$ near ${\bf k}={\bf 0}$ and are analyzed in Sec.\
\ref{sec:low_symm}.

The MPC and XC correction methods are compared in Sec.\
\ref{sec:mpc_equiv_fscorr}.  The near equivalence of the MPC and the $\Delta
V_{\rm Ew}$ correction in cubic systems is proved very directly in Appendix
\ref{app:corr_mpc_equiv}.

\subsection{Finite-size correction to the KE in 3D \label{subsec:fse-ke}}

According to the inhomogeneous
generalization\cite{malatesta_1997,gaudoin_2001,wood_2006} of the Bohm-Pines
RPA,\cite{bohm_1953} which is believed to provide an accurate description of
long-ranged correlations of electrons in solids, the wave function of a
many-electron system may be approximated as
\begin{equation}
\Psi = \Psi_s \exp \left ( \frac{1}{2\Omega} \sum_{{\bf G}_s} u({\bf G}_s)
\Delta \hat{\rho}^{\ast}({\bf G}_s)  \Delta \hat{\rho}({\bf G}_s) \right ),
\end{equation}
where $\Delta \hat{\rho} = \hat{\rho} - \rho$ and $\Psi_s$ has short-ranged
  correlations only.   Expressed in terms of the coordinate operators, the RPA
  wave function takes the familiar\cite{foulkes_2001} form
\begin{equation}
\Psi = \Psi_s \exp \left ( \frac{1}{2}\sum_{i,j} u({\bf r}_i - {\bf r}_j) +
\sum_i \chi({\bf r}_i) \right ) ,
\end{equation}
where $\chi({\bf r}) = -\int_\Omega u({\bf r} - {\bf r}^\prime) \rho({\bf
r}^\prime) \, d{\bf r}^\prime$.

The long-ranged correlations are described by the function $u({\bf r})$, which
has the periodicity of the simulation cell and inversion symmetry. At large
$r$, $u({\bf r})$ is spin-independent and, in a  3D system, usually decays
like $1/r$. However, $u({\bf r})$ is  necessarily restricted in a finite
simulation cell, thereby biasing  the KE\@.

In a VMC simulation, the KE is evaluated as the average of the sampled values
of\cite{foulkes_2001}
\begin{equation}
\hat{T}= -\frac{1}{4} \nabla^2 \log(\Psi) = \hat{T}_s-\frac{1}{8 \Omega}
\sum_{{\bf G}_s \neq {\bf 0}} u({\bf G}_s) \nabla^2 \left [ \Delta
\hat{\rho}^\ast ({\bf G}_s) \Delta \hat{\rho}({\bf G}_s) \right] ,
\end{equation}
where $\hat{T}_s=-\nabla^2 \log(\Psi_s)/4$ and $\nabla =
(\nabla_1,\ldots,\nabla_N)$ is the $3N$-dimensional gradient operator.  It can
easily be shown that $\nabla^2 \left [ \Delta \hat{\rho}^\ast({\bf G}_s)
\Delta \hat{\rho}({\bf G}_s) \right ] = - G_s^2 \hat{\rho}^\ast ({\bf G}_s)
\Delta \hat{\rho}({\bf G}_s) - G_s^2 \hat{\rho}({\bf G}_s) \Delta
\hat{\rho}^{\ast}({\bf G}_s) + 2 G_s^2 N$. Since $\langle \Delta
\hat{\rho}({\bf G}_s)\rangle = 0$, and hence $\langle \hat{\rho}({\bf G}_s)
\Delta \hat{\rho}({\bf G}_s)\rangle = \langle \Delta \hat{\rho}^{\ast}({\bf
G}_s) \Delta \hat{\rho}({\bf G}_s)\rangle$, it follows that
\begin{eqnarray}
\left< \hat{T} \right> & = & \left< \hat{T}_s \right> + \frac{1}{4\Omega}
\sum_{{\bf G}_s \neq {\bf 0}} G_s^2 \left[ u({\bf G}_s) \left< \Delta
\hat{\rho}^\ast ({\bf G}_s) \Delta \hat{\rho} ({\bf G}_s) \right> -  N u({\bf
G}_s) \right]
\label{eqn:expec_T_a} \\ & = & \left< \hat{T}_s \right> + \frac{N}{4
\Omega} \sum_{{\bf G}_s \neq {\bf 0}} G_s^2 u({\bf G}_s) S^\ast ({\bf G}_s) -
\frac{N}{4 \Omega} \sum_{{\bf G}_s\neq {\bf 0}} G_s^2 u({\bf G}_s).
\label{eqn:expec_T}
\end{eqnarray}
We assume that $\langle \hat{T}_s \rangle$ is exactly proportional to the
system size (i.e., any finite-size error in $\langle \hat{T}_s \rangle$ has
been eliminated by twist averaging or the use of HF corrections) and
concentrate here on the long-ranged finite-size errors arising from the
Jastrow factor.  Although the sum over ${\bf G}_s$ in Eq.\
(\ref{eqn:expec_T_a}) converges, the two contributing terms in Eq.\
(\ref{eqn:expec_T}) diverge. As in the analysis of the Coulomb errors in Sec.\
\ref{subsec:fscee}, this difficulty can be overcome by the inclusion of
convergence factors, which are to be understood in the rest of this work.

In practice $u({\bf k})$ has roughly the same form at different system sizes,
since its small-${\bf k}$ behavior is determined by the
RPA\@.\cite{fin_chiesa} Hence, in the infinite-system limit, the sum over
${\bf G}_s$ in Eq.\ (\ref{eqn:expec_T}) can be replaced by an integral without
changing the function $u({\bf k})$.  For a symmetric system, $u({\bf r})
=-A/r$ for large $r$, where $A$ is a constant,\cite{bohm_1953} so $u({\bf
k})=-4\pi A/k^2$ at small ${\bf k}$. Therefore $\lim_{{\bf k}\rightarrow {\bf
0}} k^2 u({\bf k})$ is finite, and the leading contribution to the finite-size
error is the omission of the ${\bf G}_s={\bf 0}$ term in the second summation
in Eq.\ (\ref{eqn:expec_T}). The ${\bf G}_s = {\bf 0}$ term in the first
summation is less important because $S({\bf k}) =O(k^2)$.  This argument leads
to the finite-size correction proposed by Chiesa \textit{et
al.}:\cite{fin_chiesa}
\begin{equation}
\Delta T = \frac{N \pi A}{\Omega}. \label{eqn:lead_order_ke}
\end{equation}
In the HEG, where the RPA implies that $A=1/\omega_p$,\cite{bohm_1953,pines}
this correction becomes $\Delta T = \omega_p/4$.

\subsection{Finite-size corrections within a DFT framework}

Kwee \textit{et al.}\ have recently proposed an approach for removing
finite-size errors from QMC data by computing a correction within
DFT\@.\cite{kwee} The correction is given by the difference between the DFT
energy evaluated using the local-density-approximation (LDA) functional which
is appropriate for an infinite system and the DFT energy evaluated with an LDA
functional modified to be appropriate for an $N$-electron system.  The
parameters for the modified LDA are obtained from DMC calculations for
$N$-electron HEGs.  The approach is successful in the examples studied by Kwee
\textit{et al.},\cite{kwee} but does not shed any light on how to correct
finite-size errors in the HEG itself.  This approach relies on the LDA (or
another density functional) being a reasonable description of the system under
study, whereas the approaches discussed in this article are not restricted in
this manner.

\section{Comparison of twist-averaging methods \label{sec:twistav_comp}}

HF theory is the simplest framework in which twist-averaging methods can be
compared. Very large simulation cells and twist samplings can be used,
allowing the convergence with cell size and number of twists to be assessed
reliably. Some of the finite-size errors that affect real interacting systems
are not present in HF calculations, but twist averaging is only intended to
remove single-particle errors and the HF framework provides a valid test of
how well it achieves this aim.

The first issue is the choice of quadrature.  The integrations over the
simulation-cell Brillouin zone that yield twist-averaged energies cannot be
carried out exactly and must be approximated by sums over finite sets of ${\bf
k}_s$-points. We have considered three choices for the set of points: (i) a
uniform Monkhorst-Pack grid\cite{monkhorst_1976} centered on the
$\Gamma$-point of the simulation-cell Brillouin zone, (ii) a uniform grid
centered on the Baldereschi-point\cite{baldereschi} of the simulation-cell
Brillouin zone, and (iii) a random sampling within the simulation-cell
Brillouin zone. All three choices yield identical results as the number of
electrons $N$ or the number of twists $M$ tends to infinity, but the two
limits are not equivalent: the fully twist-averaged ($M \rightarrow \infty$)
exchange energy depends strongly on $N$ in both ensembles, while the fully
twist-averaged KE depends weakly on $N$ in the CE and has no
systematic error in the GCE\@. Since practical QMC simulations are unlikely to
use very large simulation cells or numbers of twists (large numbers of twists
are difficult because the full many-electron trial wave function must be
constructed and stored for each twist), the rates of convergence with $N$ and
$M$ are important.

If the system is an insulator, the same bands are occupied at every ${\bf
k}_s$ and the integrand (e.g., the total KE as a function of ${\bf
k}_s$) is very smooth. The sampling theorem then ensures that estimates of the
integral obtained using uniform twist grids converge very rapidly as the
number of twists $M$ is increased.  If the twists are distributed randomly,
the statistical error in the estimate of the integral decays more slowly, like
$M^{-1/2}$.  The most rapid convergence with number of twists and system size
is obtained using a uniform grid of twists offset to the Baldereschi
point\cite{baldereschi} of the simulation-cell Brillouin zone.

\begin{figure}
\begin{center}
\includegraphics[scale=0.3,clip]{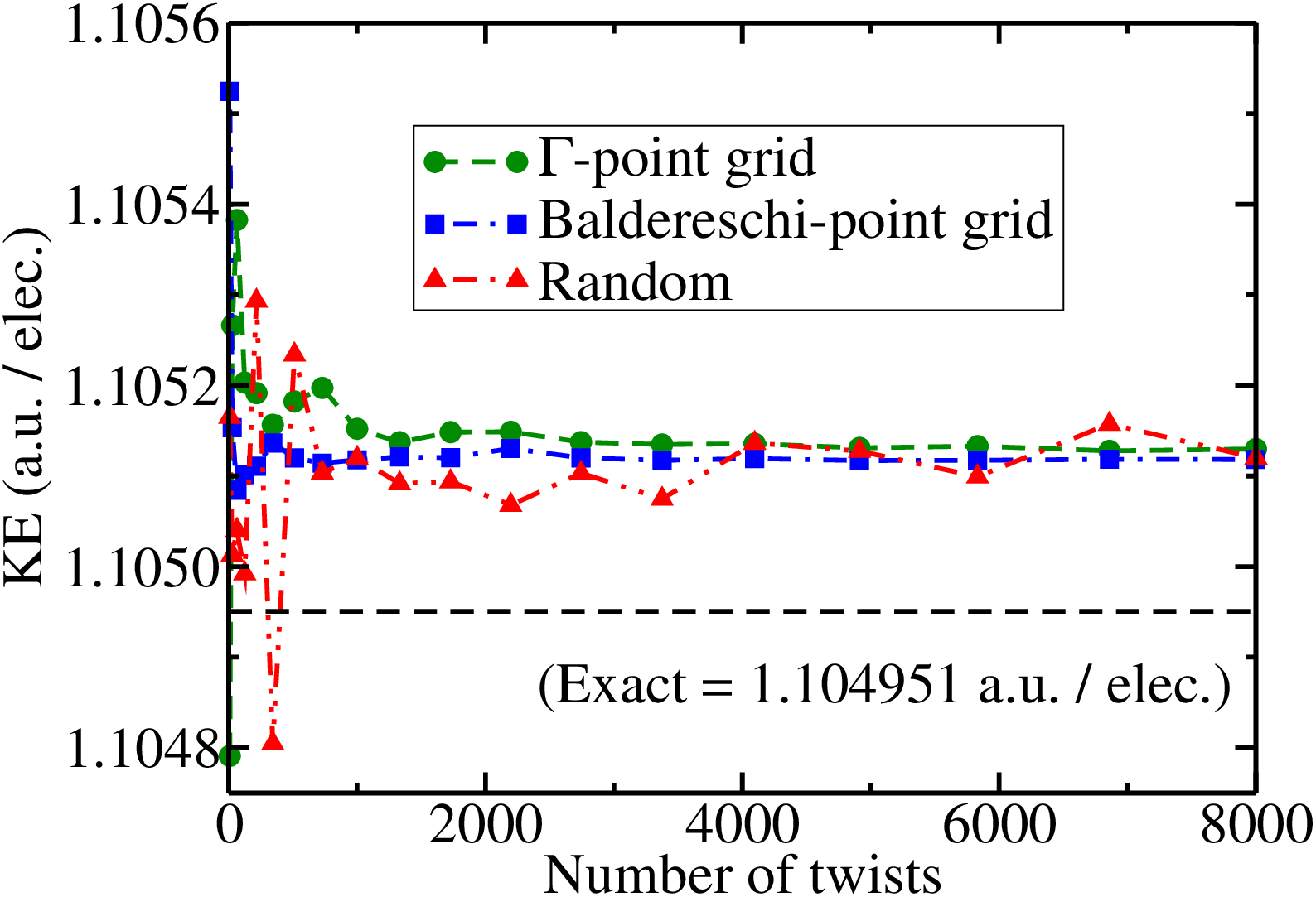} \\[1ex]
\includegraphics[scale=0.3,clip]{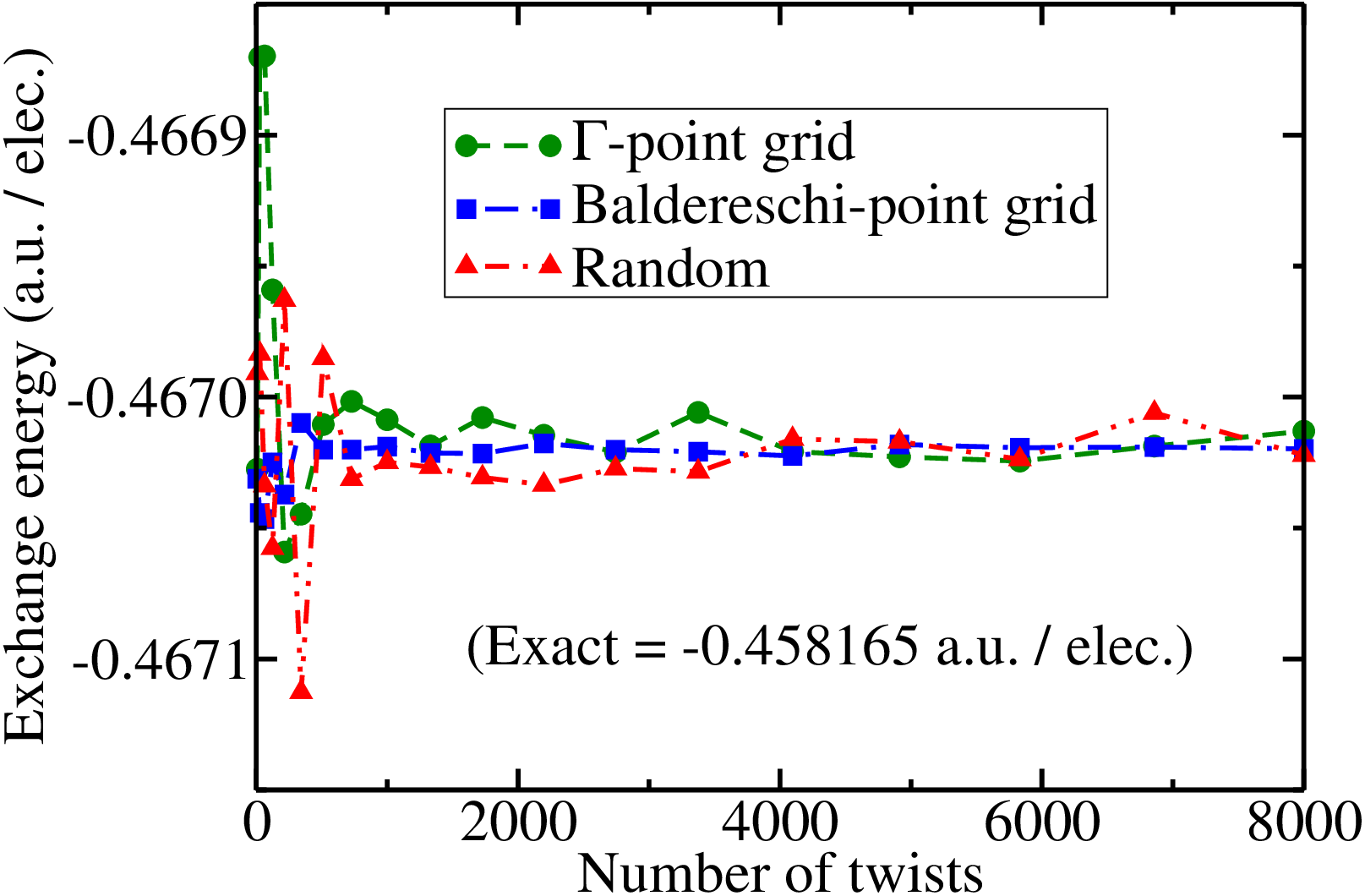} \\[1ex]
\caption{(Color online) Convergence of the calculated KE per electron (top
  panel) and exchange energy per electron (bottom panel) of a 338-electron
  simulation cell of HEG at $r_s=1$ a.u.\ as a function of the number of
  twists for the three different CE twist-averaging methods described in the
  text. Because of the finite size of the simulation cell, the calculated KE
  and exchange energy do not converge to their exact infinite-system limits as
  the number of twists increases: the KE shows a small positive bias and the
  exchange energy a large negative bias.
\label{fig:twist_convergence}}
\end{center}
\end{figure}

In metals, the integrand is discontinuous because of the sharp Fermi surface
and the convergence with system size and number of twists is much slower.
Figure \ref{fig:twist_convergence} shows the HF
kinetic and exchange energies of a face-centered cubic (FCC) simulation cell
of HEG containing 338 electrons at $r_s=1$ a.u., calculated using sets of
twists of various sizes generated in all three ways.  As for insulators,
energies calculated using random twist sampling converge slowly as the number
of twists increases.  The most rapid convergence is again obtained with a
uniform Monkhorst-Pack grid of twists centered on the Baldereschi point of the
simulation-cell Brillouin zone. The twists on a $\Gamma$-point Monkhorst-Pack
belong to stars of symmetry-equivalent twists yielding identical energies.
The symmetry can be used to reduce the number of trial wave functions that
have to be constructed, optimized and stored per twist, but does not decrease
the total number of Monte Carlo samples required to obtain a given statistical
error and does not affect the conclusion that the Baldereschi-point grid is
the most efficient. Because the simulation cell only contains 338 electrons,
the KE and exchange energy do not converge to their infinite-system limits as
the number of twists increases.  The small positive error in the calculated KE
is an artifact of the CE twist-averaging algorithm, as discussed in Sec.\
\ref{subsec:twist-averaging}, and disappears when GCE twist averaging is
used. KE's in QMC simulations suffer from much larger finite-size errors due
to long-ranged correlations (see Sec.\ \ref{subsec:fse-ke}), but these are
absent in HF theory.  The large negative finite-size error in the exchange
energy is not caused by the CE twist-averaging algorithm and is not removed by
GCE averaging, but arises from the compression of the exchange hole into the
simulation cell.

\begin{figure}
\begin{center}
\includegraphics[scale=0.3,clip]{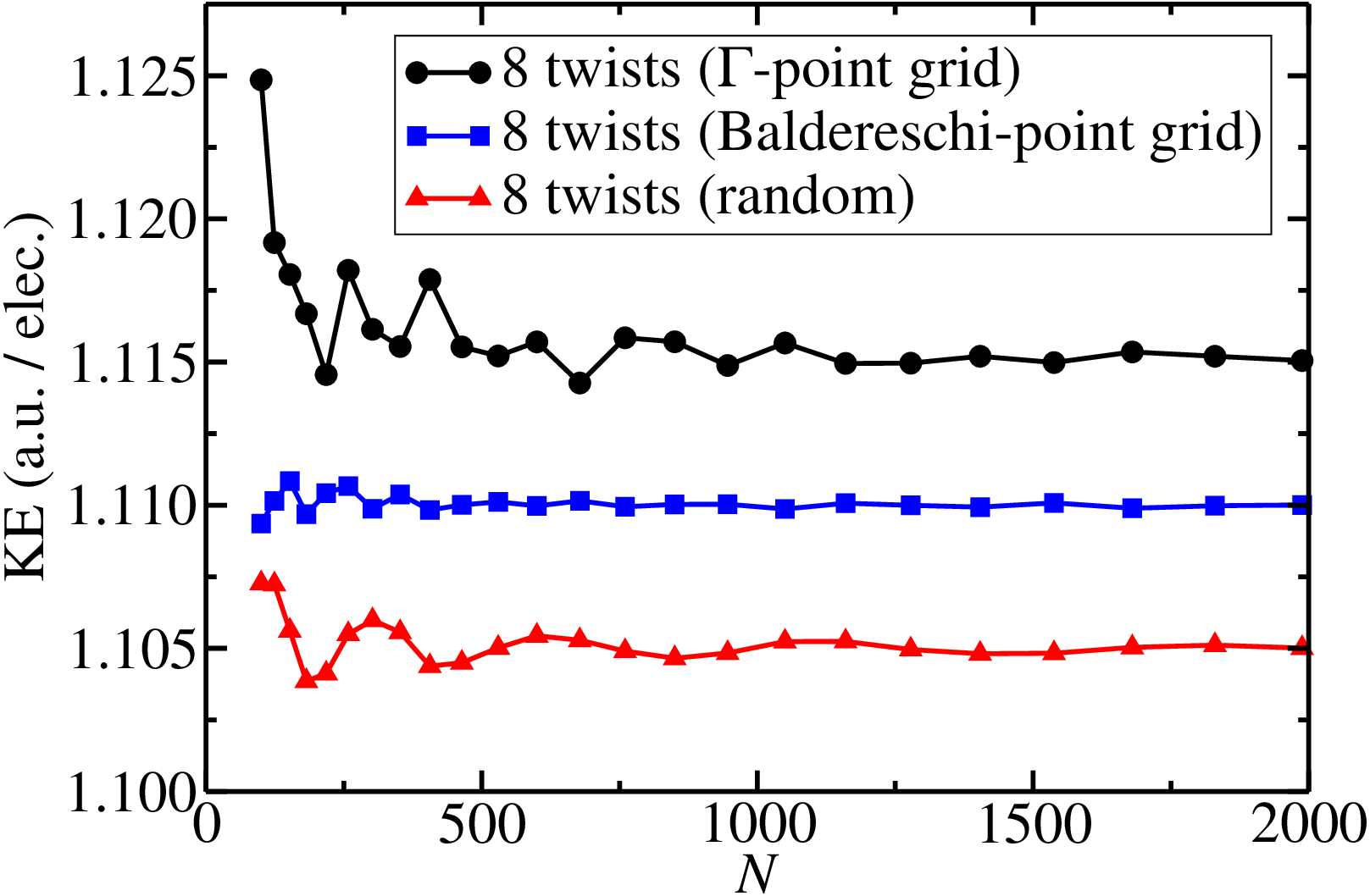}
\\[1ex]
\includegraphics[scale=0.3,clip]{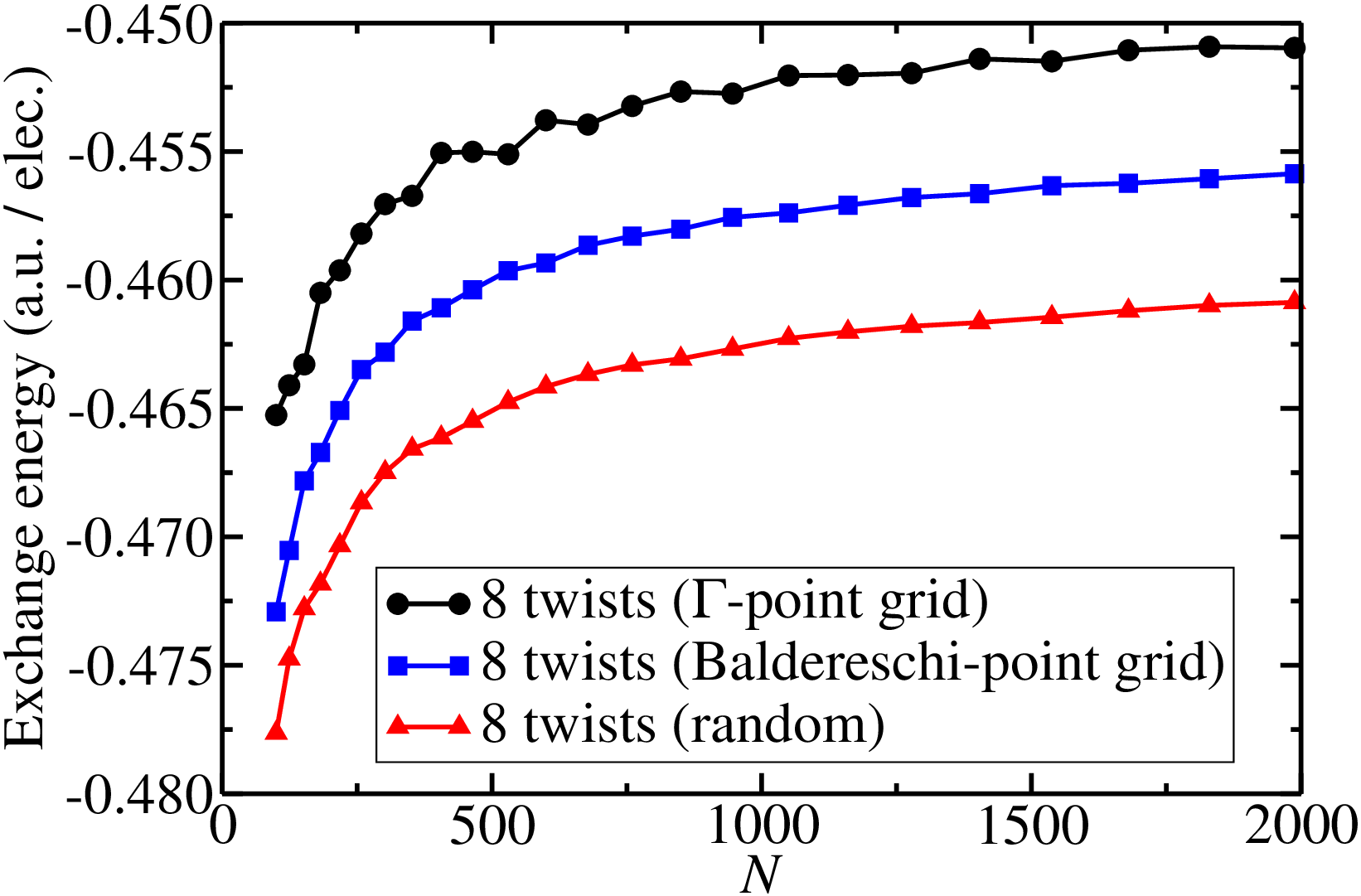}
\\[1ex]
\caption{(Color online) Convergence of the calculated KE per electron (top
  panel) and exchange energy per electron (bottom panel) of a HEG at $r_s=1$
  a.u.\ as a function of $N$ for the three different CE twist-averaging
  methods described in the text, each with eight twists.  The $\Gamma$-point
  and Baldereschi-point results have been offset for clarity.
\label{fig:twist_sampling_comparison}}
\end{center}
\end{figure}

Figure \ref{fig:twist_sampling_comparison} shows the convergence with
system size of the CE twist-averaged HF KE and exchange energies of a HEG at
$r_s=1$ a.u.\ in an FCC simulation cell, calculated using
sets of twists generated in all three ways. To highlight the differences
between the three methods, we have used only eight twists in each case.
Energies calculated using the uniform grid of twists centered on $\Gamma$
converge the most slowly because of the large fluctuations that occur as the
size of the simulation cell increases and shells of symmetry-equivalent ${\bf
k}_s + {\bf G}_s$ vectors cross the Fermi surface. Energies calculated using a
random sampling of twists converge more rapidly with system size (although
less rapidly with number of twists). Yet again, the best approach uses a
uniform grid of twists centered on the Baldereschi point of the
simulation-cell Brillouin zone.

\begin{figure}
\begin{center}
\includegraphics[scale=0.3,clip]{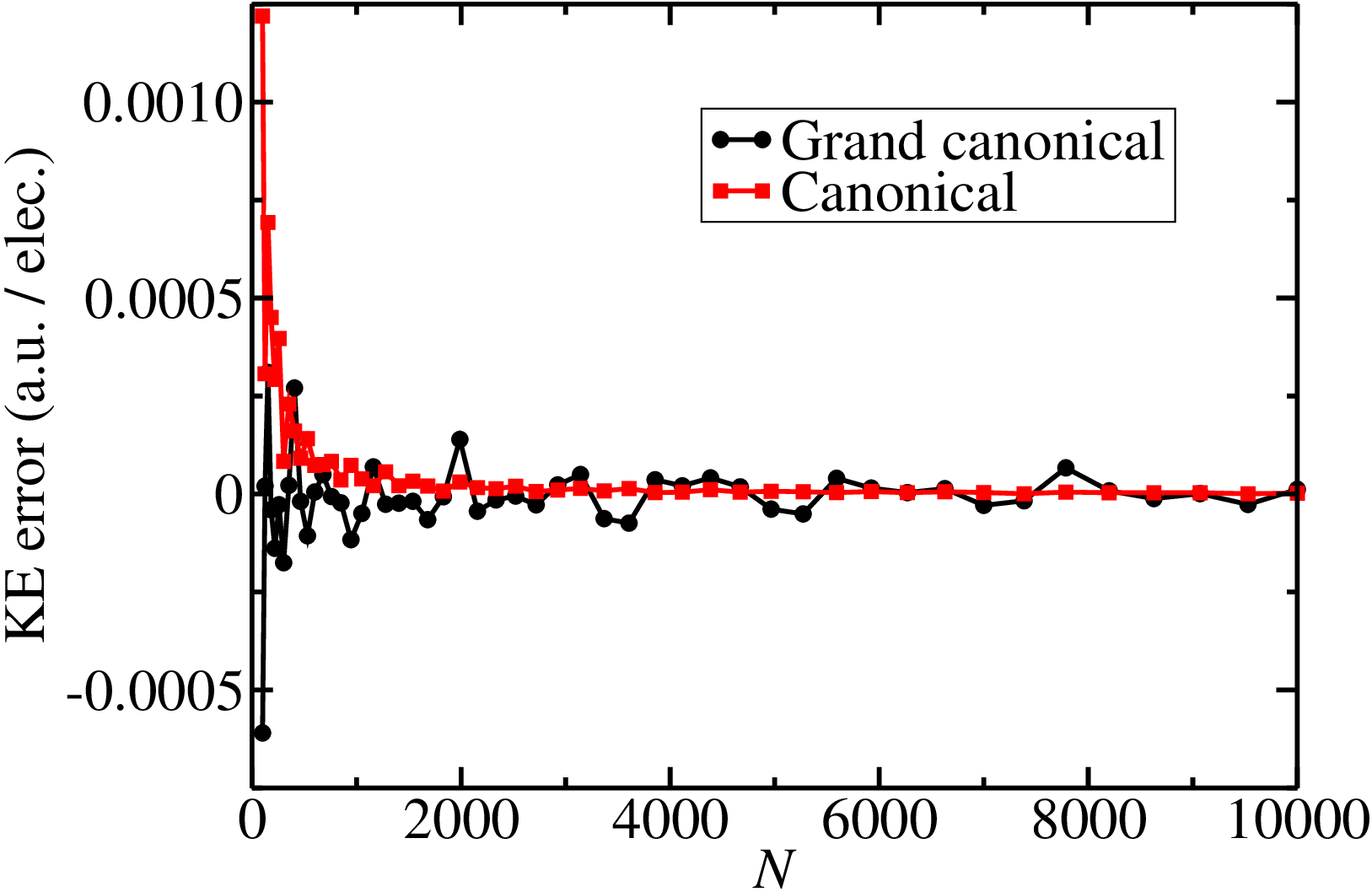} \\[1ex]
\caption{(Color online) System-size dependence of the twist-averaged KE per
  electron of a HEG at $r_s = 1$ a.u.\ in the CE and GCE\@. All calculations
  used a random sampling of 5120
  twists. \label{fig:canonical_vs_grand_canonical}}
\end{center}
\end{figure}

\begin{figure}
\begin{center}
\includegraphics[scale=0.3,clip]{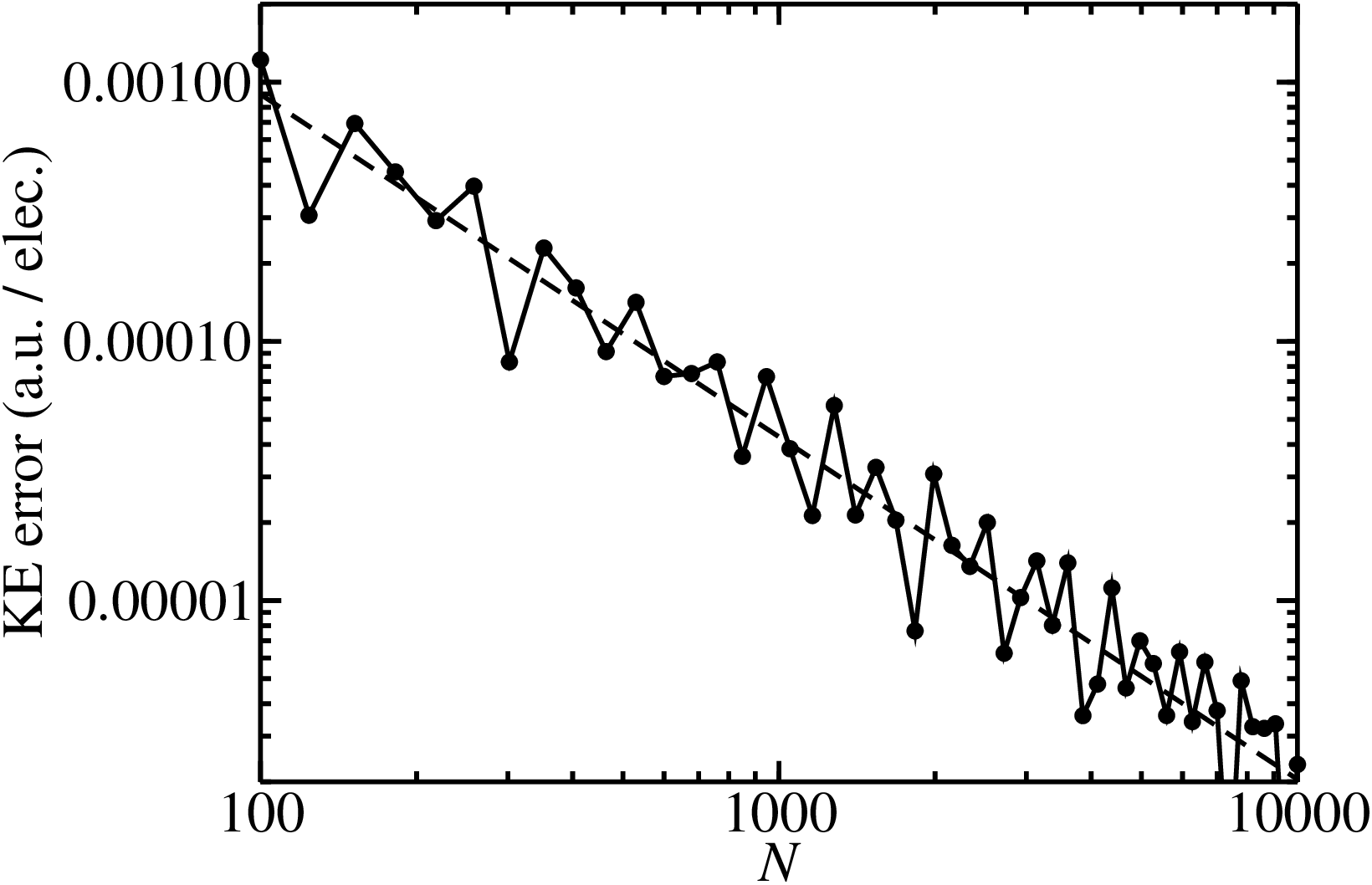}
\\[1ex]
\caption{Bias in the KE per electron of a HEG at $r_s=1$ a.u.\ as a function
  of $N$, calculated using 5120 randomly chosen twists in the CE\@. The
  power-law fit yields a bias proportional to
  $N^{-1.32}$. \label{fig:canonical_error}}
\end{center}
\end{figure}

Figure \ref{fig:canonical_vs_grand_canonical} shows the error in the
twist-averaged HF KE calculated with a very large set of random twists, using
both the CE and GCE\@. The systematic bias in the CE average disappears when
GCE averaging is used, but the large fluctuations in the GCE results outweigh
the bias for all but the smallest simulation cells. These fluctuations arise
from the variations in electron number inherent in the GCE method. Most QMC
simulations are likely to use many fewer twists, rendering the GCE
fluctuations even worse, so CE averaging is the more promising method despite
the bias.  Figure \ref{fig:canonical_error} shows the bias in the
CE-twist-averaged
KE as a function of $N$. The power-law fit shows that the bias per electron
decreases relatively slowly with system size, scaling roughly as $N^{-4/3}$,
as noted by Lin \textit{et al.}\cite{lin_twist_2001}

\section{Comparison of the MPC interaction with the finite-size
correction to the Ewald energy \label{sec:mpc_equiv_fscorr}}

If the XC hole can be assumed to have converged to its infinite-system form
then both the MPC interaction and the finite-size correction of Eq.\
(\ref{eq:finsize1}) are good solutions to the problem of finite-size effects
in the XC energy of a cubic system.  For low-symmetry systems the MPC
interaction should continue to be a good solution, whereas the correction to
the Ewald energy cannot be applied straightforwardly.  On the other hand, if
the simulation cell is too small to contain the infinite-system XC hole, but
the SF is known analytically at small $k$, then this information  can be
included in the XC correction but not the MPC interaction, so the XC
correction may work better.  In practice the difference between the MPC energy
and the corrected Ewald energy for cubic interacting systems is very small
when the Ewald interaction is used to generate the configuration distribution,
as demonstrated by the data shown for 3D HEGs at three different densities in
Table \ref{table:heg_ewald_v_mpc}.  In each case the difference of MPC and
Ewald energies is approximately equal to (but slightly greater than) $\Delta
V_{\rm Ew}$.

\begin{table}
\begin{center}
\begin{tabular}{ccr@{.}lr@{.}lc}

\hline \hline

$r_s$ & $N$ & \multicolumn{2}{c}{$(E_{\rm MPC}-E_{\rm Ewald})/N$} (a.u.\ /
elec.) & \multicolumn{2}{c}{$\Delta V_{\rm Ew}/N$ (a.u.\ / elec.)} & \%age
difference \\

\hline

1  &  54 & ~~~~~~$0$&$007\,81(1)$   & ~~~~$0$&$008\,02$   & 2.6(1)\% \\

1  & 102 &       $0$&$004\,137(9)$  &     $0$&$004\,245$  & 2.5(2)\% \\

1  & 226 &       $0$&$001\,89(1)$   &     $0$&$001\,92$   & 1.6(5)\% \\

3  &  54 &       $0$&$001\,551(4)$  &     $0$&$001\,543$  & 0.5(3)\% \\

3  & 102 &       $0$&$000\,802(2)$  &     $0$&$000\,817$  & 1.8(2)\% \\

3  & 226 &       $0$&$000\,365(1)$  &     $0$&$000\,369$  & 1.1(3)\% \\

10 &  54 &       $0$&$000\,242(1)$  &     $0$&$000\,254$  & 4.7(4)\% \\

10 & 102 &       $0$&$000\,131\,9(4)$ &   $0$&$000\,134\,2$ & 1.7(3)\% \\

10 & 226 &       $0$&$000\,060\,5(7)$ &   $0$&$000\,060\,6$ & 0(1)\% \\

\hline \hline
\end{tabular}
\caption{Difference between total energies per electron evaluated using the
MPC and Ewald interactions in twist-averaged DMC calculations for 3D
paramagnetic HEGs at three different densities.  The Ewald energy is used in
the branching factor in the DMC simulation, so that the configuration
distribution appropriate for the Ewald interaction is used in all cases.  The
DMC time steps were 0.003, 0.03, and 0.3 a.u.\ at $r_s=1$, 3, and 10 a.u.,
respectively, and the target population was more than 400 configurations in
each case.  Twist angles were sampled randomly.  At each density it was
verified that the DMC energy did not change when the time step was halved, the
configuration population was doubled, and the number of post-twist-change
equilibration steps was quadrupled.  The finite-size correction to the Ewald
energy [Eq.\ (\ref{eqn:xc_corr_heg})] is shown for comparison.
\label{table:heg_ewald_v_mpc}}
\end{center}
\end{table}

It is shown in Appendix \ref{sec:hf_fs_errors} that the long range of the
exchange hole causes the MPC energy to be slowly convergent when the
interactions are treated within the HF approximation. The finite-size
correction constructed using the known small-$k$ behavior of the HF SF
therefore performs better than the MPC interaction in HF calculations.

\begin{figure}
\begin{center}
\includegraphics[scale=0.3,clip]{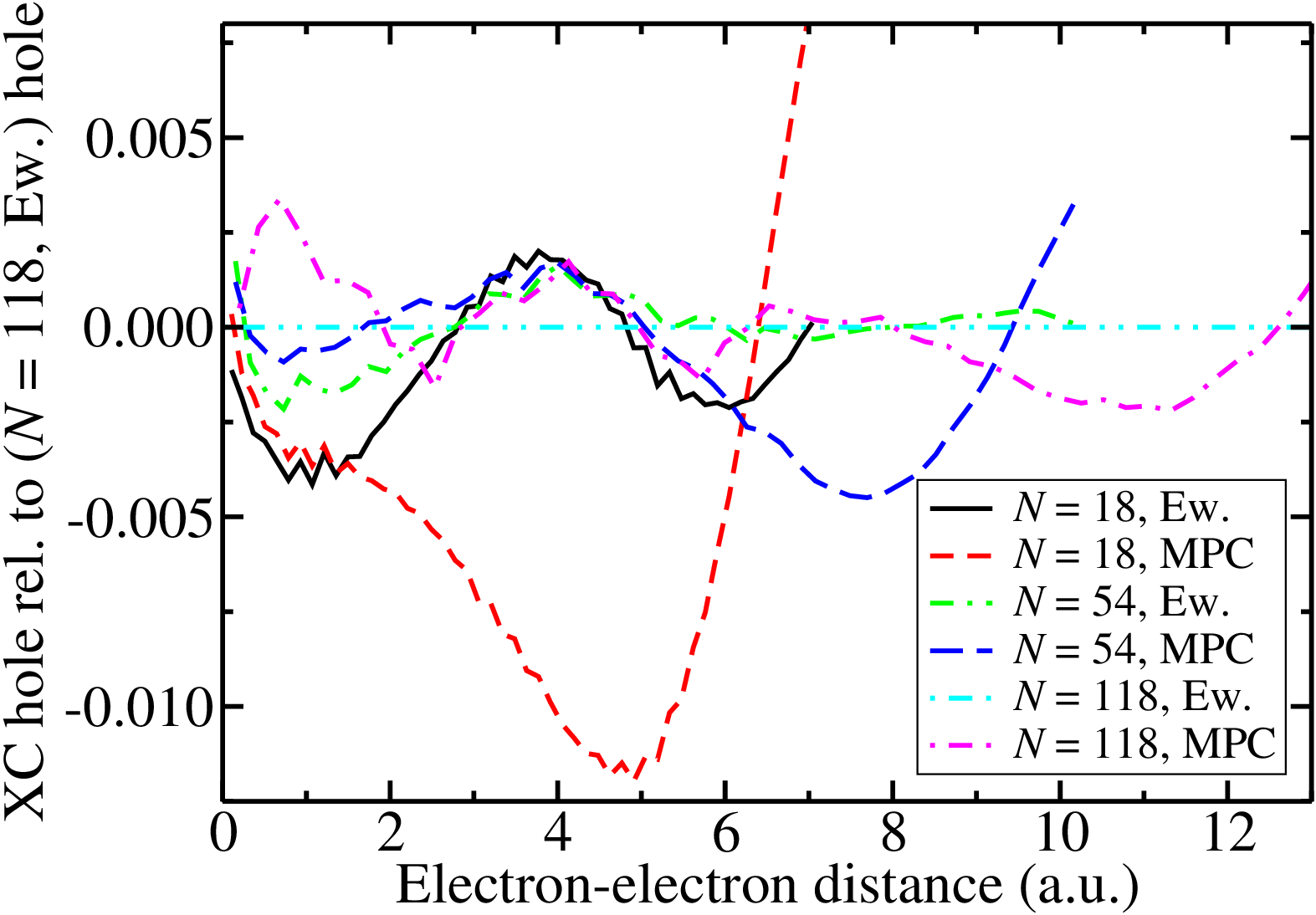}
\caption{(Color online) System-averaged XC hole in a 3D paramagnetic HEG of
  density parameter $r_s=3$\ a.u.\ at different system sizes relative to the
  XC hole in a 118-electron HEG with the Ewald interaction. The Ewald and MPC
  interactions were used to generate the configuration distributions.
  Twist-averaged VMC and DMC XC holes $\rho_{\rm xc}^{\rm VMC}(r)$ and
  $\rho_{\rm xc}^{\rm DMC}(r)$ were calculated, and the final XC hole was
  obtained using the extrapolated estimate $\rho_{\rm xc}(r) \approx 2
  \rho_{\rm xc}^{\rm DMC}(r)-\rho_{\rm xc}^{\rm VMC}(r)$.  The error in the
  extrapolated estimate is second order in the error in the trial wave
  function.\cite{foulkes_2001}
\label{fig:Para_rs3_XChole}}
\end{center}
\end{figure}

By the variational principle, the expectation value of the MPC Hamiltonian
with respect to the Ewald ground-state wave function is greater than the
expectation value of the MPC Hamiltonian with respect to the MPC ground-state
wave function. The MPC energy obtained using DMC with the Ewald energy in the
branching factor is therefore likely to be overestimated, and \textit{vice
versa}.  An example of this effect is shown in Table
\ref{table:heg_ewald_v_mpc_propagation}.  When the Ewald interaction is used
in the branching factor, the difference between the MPC and Ewald energies is
given by $\Delta V_{\rm Ew}$.  However, when the MPC interaction is used, the
difference is less than $\Delta V_{\rm Ew}$.  These results suggest that the
MPC interaction distorts the XC hole in a finite system, while the Ewald
interaction gives a better shaped hole, although the interaction with the hole
is not quite right.  We have directly verified that this is the case for a HEG
at $r_s=3$\ a.u., as can be seen in Fig.\ \ref{fig:Para_rs3_XChole}.  The
Ewald XC hole converges to its infinite-system form much more rapidly than the
MPC hole.  The likely reason for this behavior is that the MPC Hamiltonian
does not include corrections for finite-size errors in the KE\@.

\begin{table}
\begin{center}
\begin{tabular}{cr@{.}lr@{.}lr@{.}lr@{.}l}

\hline \hline

    & \multicolumn{4}{c}{Ewald propagation} & \multicolumn{4}{c}{MPC
    propagation} \\

\raisebox{1ex}[0pt]{$N$} & \multicolumn{2}{c}{$E_{\rm Ew}$}/N (a.u.\ / elec.)
& \multicolumn{2}{c}{$E_{\rm MPC}/N$ (a.u.\ / elec.)} &
\multicolumn{2}{c}{$E_{\rm Ew}/N$} (a.u.\ / elec.) &
\multicolumn{2}{c}{$E_{\rm MPC}/N$ (a.u.\ / elec.)} \\

\hline

54  & ~~~~~$-0$&$068\,69(6)$ & ~~~~~$-0$&$067\,15(6)$ & ~~~~~$-0$&$068\,18(6)$
& ~~~~~$-0$&$067\,36(6)$ \\

102 & $-0$&$067\,62(3)$ & $-0$&$066\,82(3)$ & $-0$&$067\,68(6)$ &
$-0$&$067\,15(6)$ \\

226 & $-0$&$067\,06(4)$ & $-0$&$066\,61(4)$ & $-0$&$066\,85(5)$ &
$-0$&$066\,77(5)$ \\

\hline \hline
\end{tabular}
\caption{Total energies evaluated using Ewald and MPC interactions for 3D
paramagnetic HEGs at $r_s=3$ a.u.  The results were obtained in twist-averaged
DMC calculations, as described in the caption to Table
\ref{table:heg_ewald_v_mpc}. The Ewald energy was used in the branching factor
in the results labeled ``Ewald propagation,'' while the MPC energy was used in
the results labeled ``MPC propagation'' (i.e., the XC hole was appropriate for
the Ewald and MPC interactions, respectively).  $E_{\rm Ew}$ and $E_{\rm MPC}$
refer to the interaction (Ewald and MPC, respectively) used in the local
energies that were averaged to obtain the DMC energy.
\label{table:heg_ewald_v_mpc_propagation}}
\end{center}
\end{table}

\section{Nonanalytic behavior at ${\bf k}={\bf 0}$
  \label{sec:nonanalytic}}

\subsection{Examples of nonanalytic behavior at ${\bf k}={\bf 0}$
\label{sec:ex_na}}

The XC correction discussed in Sec.\ \ref{subsec:fscee} works well for
interacting systems of cubic symmetry.  In other cases, however, the theory
cannot be applied straightforwardly.  We give two examples.

For a general interacting system, the SF at small $k$ can be written as
$S({\bf k})=(1/2) {\bf k}^T W^\prime {\bf k}$ for some tensor $W^\prime$.  If
the system has cubic symmetry then $W^\prime$ is proportional to the identity
matrix and $\lim_{{\bf k}\rightarrow {\bf 0}} S({\bf k})/k^2$ is well-defined.
Otherwise, this limit is undefined and it is not possible to add the ${\bf
G}_s={\bf 0}$ term to the sum in Eq.\ (\ref{eq:dvew2}).

Within HF theory, $S({\bf k})= \lambda k + O(k^3)$ at small $k$, where
$\lambda$ is a constant.\cite{giuliani} The limit of $S({\bf k})/k^2$ as ${\bf
k}\rightarrow {\bf 0}$ is therefore undefined.  Again the approach discussed
in Sec.\ \ref{subsec:fscee} cannot be applied.

\subsection{Removing the problematic part of the SF \label{sec:sol_na}}

Suppose that $S({\bf k})/k^2$ is singular or otherwise ill-defined at ${\bf
k}={\bf 0}$, but that its small-${\bf k}$ behavior is known and is roughly
independent of $N$. We can then introduce a model ``structure factor''
$S_b({\bf k})$ that incorporates the nonanalytic behavior and define $S_a({\bf
k})=S({\bf k})-S_b({\bf k})$, so that $\lim_{{\bf k} \rightarrow {\bf 0}}
S_a({\bf k})/k^2$ is well defined.  Starting from Eq.\ (\ref{eq:dvew2}) and
applying the Poisson summation formula\cite{footnote:poisson} to terms
involving $S_a$ only yields
\begin{equation}
\Delta V_{\rm Ew} = \frac{N}{2} \left( \frac{4\pi}{\Omega} \lim_{{\bf
k}\rightarrow {\bf 0}} \frac{S_a({\bf k})}{k^2} + \frac{1}{(2\pi)^3}
\int_{k<\infty} v_E(k) S_b({\bf k}) \, d{\bf k} - \frac{1}{\Omega} \sum_{{\bf
G}_s\neq {\bf 0}} v_E(G_s) S_b({\bf G}_s) \right) - \frac{N}{2}\sum_{{\bf R}_s
\neq {\bf 0}} \int_{r^\prime<\infty} \frac{S_a({\bf r}^\prime)}{|{\bf
R}_s-{\bf r}^\prime|} \, d{\bf r}^\prime, \label{eqn:isolation}
\end{equation}
where $S_a({\bf r})$ is a localized charge distribution analogous to $S_{\rm
loc}({\bf r})$ and all convergence factors have been omitted.  Since the ${\bf
k} \rightarrow {\bf 0}$ behavior of $S_a({\bf k})/k^2$ is known, and provided
that $S_b({\bf k})$ has a simple enough form, all three terms within the large
parentheses in Eq.\ (\ref{eqn:isolation}) can be evaluated straightforwardly.
Moreover, since $S_a({\bf k})$ is well-behaved as ${\bf k} \rightarrow {\bf
0}$, $S_a({\bf r})$ lacks the long-ranged tail present in $S_{\rm loc}({\bf
r})$; the summation in the final term on the right-hand side of Eq.\
(\ref{eqn:isolation}) therefore converges rapidly and should be small.  This
term is omitted from the approximate expressions for the finite-size
correction obtained below, and therefore represents the error in these
approximations.

The finite-size correction obtained by evaluating all except the final term on
the right-hand side of Eq.\ (\ref{eqn:isolation}) is accurate when $S_{a}({\bf
k}) = S({\bf k}) - S_b({\bf k})$ is smooth, implying that $S_a({\bf r})$ is
short ranged. The model structure factor $S_b({\bf k})$ should therefore match
the nonanalytic behavior of $S({\bf k})$ as closely as possible. It is also
sensible, although less important, to ensure that $S({\bf k}) - S_b({\bf k})$
is small. In practice, although $S({\bf k}) \rightarrow 1$ as $k \rightarrow
\infty$, the correction is most easily evaluated if $S_b({\bf k}) \rightarrow
0$ as $k \rightarrow \infty$.  A natural way of accomplishing this is to
include a Gaussian function $\exp(-\alpha k^2)$ as a factor. The parameter
$\alpha$ should be small enough that the Gaussian changes little on the scale
of the Fermi wave vector. In fact, although the reciprocal space summation and
integration diverge in the $\alpha\rightarrow 0$ limit, their difference
converges rapidly. One can therefore maximize the smoothness of $S_a({\bf k})$
by decreasing $\alpha$ until the calculated value of the correction has
converged.

A plausible alternative method\cite{fin_chiesa} for dealing with leading-order
nonanalyticities in $S({\bf k})/k^2$ at ${\bf k}={\bf 0}$ is to replace the
missing ${\bf G}_s={\bf 0}$ term in the sum over ${\bf G}_s$ in Eq.\
(\ref{eqn:ks_har_p_xc}) with an integral of $v_E(k)S({\bf k})$ over a sphere
of volume $(2\pi)^3/\Omega$.  This approach may be cast into the framework
discussed above by choosing $S_b({\bf k}) = S({\bf k}) \Theta(Q - k)$, where
$Q$ is the radius of the sphere of volume $(2\pi)^3/\Omega$ and $\Theta(Q-k)$
is a Heaviside step function. The function $S_a({\bf k}) = S({\bf k}) -
S_b({\bf k})$ is then zero at the origin, so the first term inside the
parentheses in Eq.\ (\ref{eqn:isolation}) vanishes. Unless the lattice is very
asymmetric, $S_b({\bf G}_s)$ is zero for all nonzero ${\bf G}_s$, and the
third term inside the parentheses in Eq.\ (\ref{eqn:isolation}) also
vanishes. Hence
\begin{equation}
\Delta V_{\rm Ew} = \frac{N}{2} \left( \frac{1}{(2\pi)^3} \int_{k < Q} v_E(k)
S({\bf k}) \, d{\bf k}  \right) - \frac{N}{2}\sum_{{\bf R}_s \neq {\bf 0}}
\int_{r^\prime<\infty} \frac{S_a({\bf r}^\prime)}{|{\bf R}_s-{\bf r}^\prime|}
\, d{\bf r}^\prime.
\label{eqn:Heaviside_correction}
\end{equation}
In this case, however, the sharp cutoff in $S_b({\bf k})$ leads to slowly
decaying oscillations in $S_b({\bf r})$ and therefore $S_a({\bf r})$. These
oscillations fall off as $r^{-2}$ and can never be regarded as negligible.
Unless $S({\bf k})/k^2$ is constant for $k<Q$, in which case this correction
is accurate by construction, the neglected real-space term in Eq.\
(\ref{eqn:Heaviside_correction}) is of the same order as the correction itself.

\subsection{Finite-size corrections in HF theory \label{subsec:simpleHF}}

Suppose $S({\bf k})=\lambda k+O(k^3)$, as is the case for systems of cubic
symmetry in HF theory.  The divergence of $S({\bf k})/k^2$ as ${\bf k}
\rightarrow {\bf 0}$ prevents Eqs.\ (\ref{eq:finsize1}) and
(\ref{eqn:general_xc_corr}) from being used to obtain finite-size corrections.
Let $S_b({\bf k})=\lambda k \exp(-\alpha k^2)$.  Working in the $\alpha
\rightarrow 0$ limit, Eq.\ (\ref{eqn:isolation})  becomes
\begin{equation}
\Delta V_{\rm Ew} = \lim_{\alpha \rightarrow 0} \frac{N}{2}  \left(
\frac{\lambda}{\pi \alpha} - \frac{4 \pi \lambda}{\Omega} \sum_{{\bf G}_s\neq
{\bf 0}} \frac{\exp \left( -\alpha G_s^2 \right)}{G_s} \right) + O(N^{-1/3}) =
\frac{C_{\rm HF} \lambda N}{\Omega^{2/3}} + O(N^{-1/3}),
\label{eqn:gen_hf_corr}
\end{equation}
where $C_{\rm HF}=2.8884$, $2.8372$, and $2.8882$ for FCC, simple cubic (SC),
and body-centered cubic (BCC) simulation cells,
respectively,\cite{footnote:program} and we have noted that the $O(k^3)$ term
in $S_a({\bf k})=S({\bf k})-S_b({\bf k})$ causes $S_a({\bf r})$ to fall off as
$r^{-6}$, giving the $O(N^{-1/3})$ correction.  For a 3D paramagnetic
HEG,\cite{giuliani} $\lambda=(3/4) [\Omega/(3N\pi^2) ]^{1/3}$, so
\begin{equation}
\Delta V_{\rm HF} = \frac{3C_{\rm HF}}{4 \pi r_s} \left(
\frac{N}{4}\right)^{1/3} + O(N^{-1/3}).
\label{eqn:hf_corr_X_B}
\end{equation}

An alternative real-space treatment of HF finite-size errors can be found in
Appendix \ref{sec:hf_fs_errors}. As shown in Fig.\ \ref{fig:HF_X_energy}, both
the real- and reciprocal-space approaches account for most of the HF Coulomb
finite-size error, although the reciprocal-space approach performs better
because it completely removes the $O(N^{1/3})$ error.

\begin{figure}
\begin{center}
\includegraphics[scale=0.3,clip]{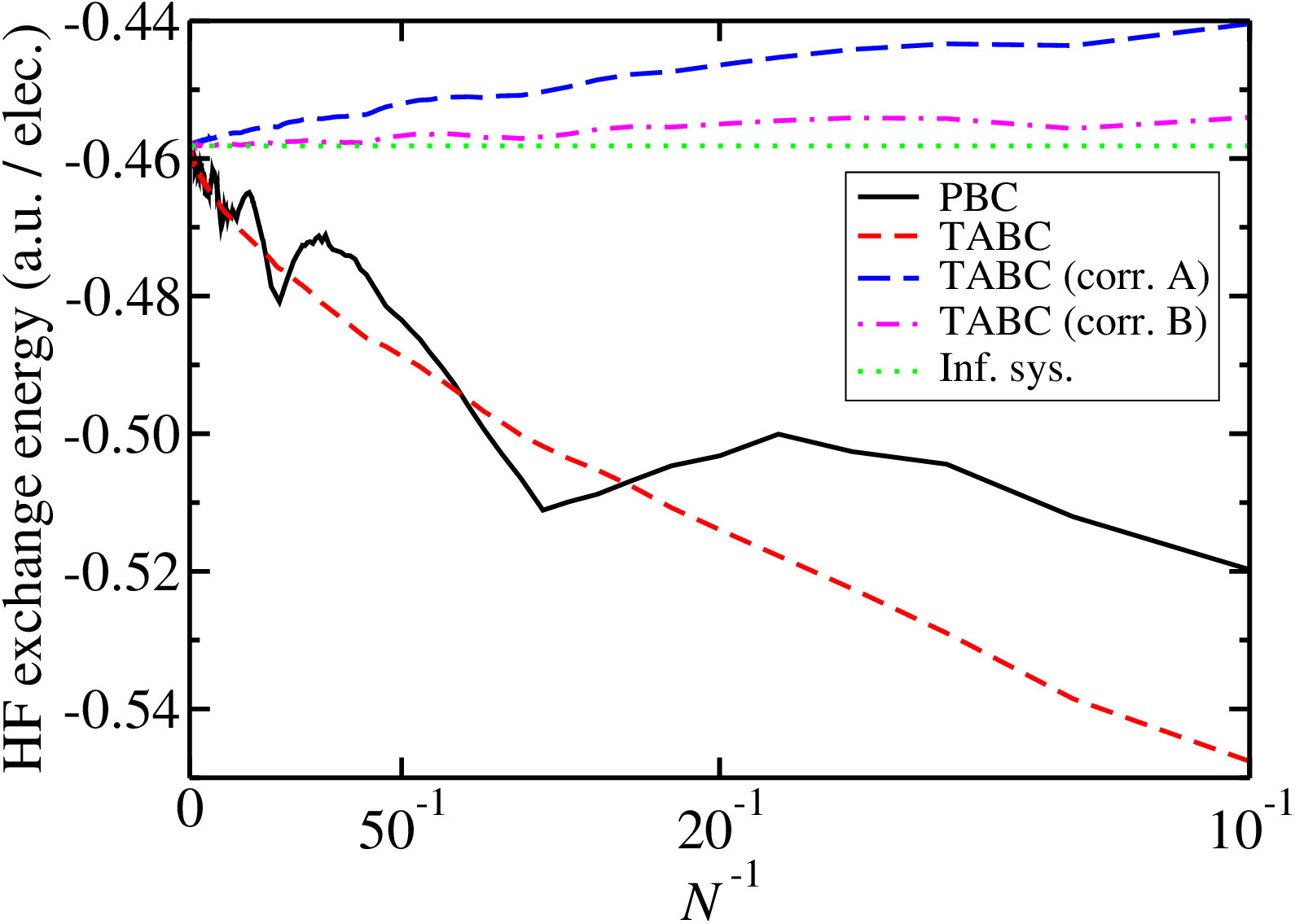}
\caption{(Color online) HF Ewald exchange energy per electron of a 3D
  paramagnetic HEG of density parameter $r_s=1$\ a.u.\ against particle number
  $N$ with ${\bf k}_s={\bf 0}$ (``PBC'') and twist-averaged boundary
  conditions within the CE (``TABC'')\@.  The corrections of Eqs.\
  (\ref{eqn:hf_corr_X_A}) and (\ref{eqn:hf_corr_X_B}) have been applied to the
  data labeled ``TABC (corr.\ A)'' and ``TABC (corr.\ B),'' respectively. An
  FCC simulation cell is used. The uncertainty in the twist-averaged data due
  to the use of a finite number of twist angles is small compared with the
  difference between the twist-averaged data and the infinite-system
  result. \label{fig:HF_X_energy}}
\end{center}
\end{figure}

\subsection{Finite-size errors in the XC energy of low-symmetry
  systems \label{sec:low_symm}}

For a general interacting system the SF can be written as
\begin{equation}
S({\bf k}) = \sum_{{\rm even}~l} \,\, \sum_{m=-l}^l S_{lm}(k) k^l
Y_{lm}(\theta_{\bf k},\phi_{\bf k}), \end{equation} where $\theta_{\bf k}$ and
$\phi_{\bf k}$ are the polar and azimuthal angles of ${\bf k}$ and $Y_{lm}$ is
the $(l,m)$-th spherical harmonic.  The odd-$l$ components are zero by
inversion symmetry.  Guided by the RPA, we assume that $S({\bf k})$ is
quadratic near ${\bf k} = {\bf 0}$, and hence that $S_{00}(k) \propto k^2$.
If the quadratic form is nonspherical, however, $l=2$ components are also
present and $\lim_{{\bf k}\rightarrow{\bf 0}}S({\bf k})/k^2$ depends on the
direction in which the limit is taken; there is then a point discontinuity at
${\bf k}={\bf 0}$.  Equivalently, the $l=2$ component gives rise to the
quadrupole moment in $S_{\rm loc}({\bf r})$, which leads to the additional
errors discussed in Sec.\ \ref{subsec:fscee}.

Let
\begin{equation} S_b({\bf k})= \sum_{m=-2}^2 S_{2m}(0) k^2
Y_{2m}(\theta_{\bf k},\phi_{\bf k}) \exp \left(-\alpha k^2 \right),
\end{equation} and $S_a({\bf k})=S({\bf k})-S_b({\bf k})$, where
$\alpha$ is such that $S_b({\bf k})$ is long-ranged in ${\bf k}$-space
compared with the Fermi wave vector.  Applying Eq.\ (\ref{eqn:isolation}) and
taking the limit $\alpha\rightarrow 0$, we find that
\begin{equation} \Delta V_{\rm Ew} = \frac{N}{2} \left( \frac{4 \pi
Y_{00}}{\Omega} \lim_{k\rightarrow 0} \frac{S_{00}(k)}{k^2} - \frac{4
    \pi}{\Omega} \sum_{m=-2}^2 S_{2m}(0) \sum_{{\bf G}_s\neq {\bf 0}}
    Y_{2m}(\theta_{{\bf G}_s},\phi_{{\bf G}_s}) \right) +
    O(N^{-1/3}). \label{eqn:lowsym_corr}
\end{equation}
In particular, it can be seen that the $O(N^0)$ finite-size correction
obtained using the spherically averaged SF is incomplete, and that there is in
general another correction of $O(N^0)$ due to the low symmetry of the
simulation cell and the existence of the $l=2$ component.  If the XC hole has
spherical symmetry, the extra correction is zero regardless of the shape of
the simulation cell; if the XC hole does not have spherical symmetry, but the
simulation cell does have cubic symmetry, the extra correction is again zero.
Hence, if one is simulating a low-symmetry system, it is advisable to choose a
simulation cell that is as close to cubic as possible.  If this is not
possible then one could evaluate the $l=0$ and $l=2$ components of $S({\bf
k})$ at ${\bf k}={\bf 0}$, and use Eq.\ (\ref{eqn:lowsym_corr}) to compute the
correction.  The $O(N^{-1/3})$ error in Eq.\ (\ref{eqn:lowsym_corr}) arises
from an assumed nonanalytic $O(k^3)$ term in $S_a({\bf r})$.

\section{Higher-order corrections to the KE \label{sec:get_ke_right}}

\subsection{Need to include higher-order corrections
  \label{sec:need_for_ke_corr}}

\begin{figure}
\begin{center}
\includegraphics[scale=0.3,clip]{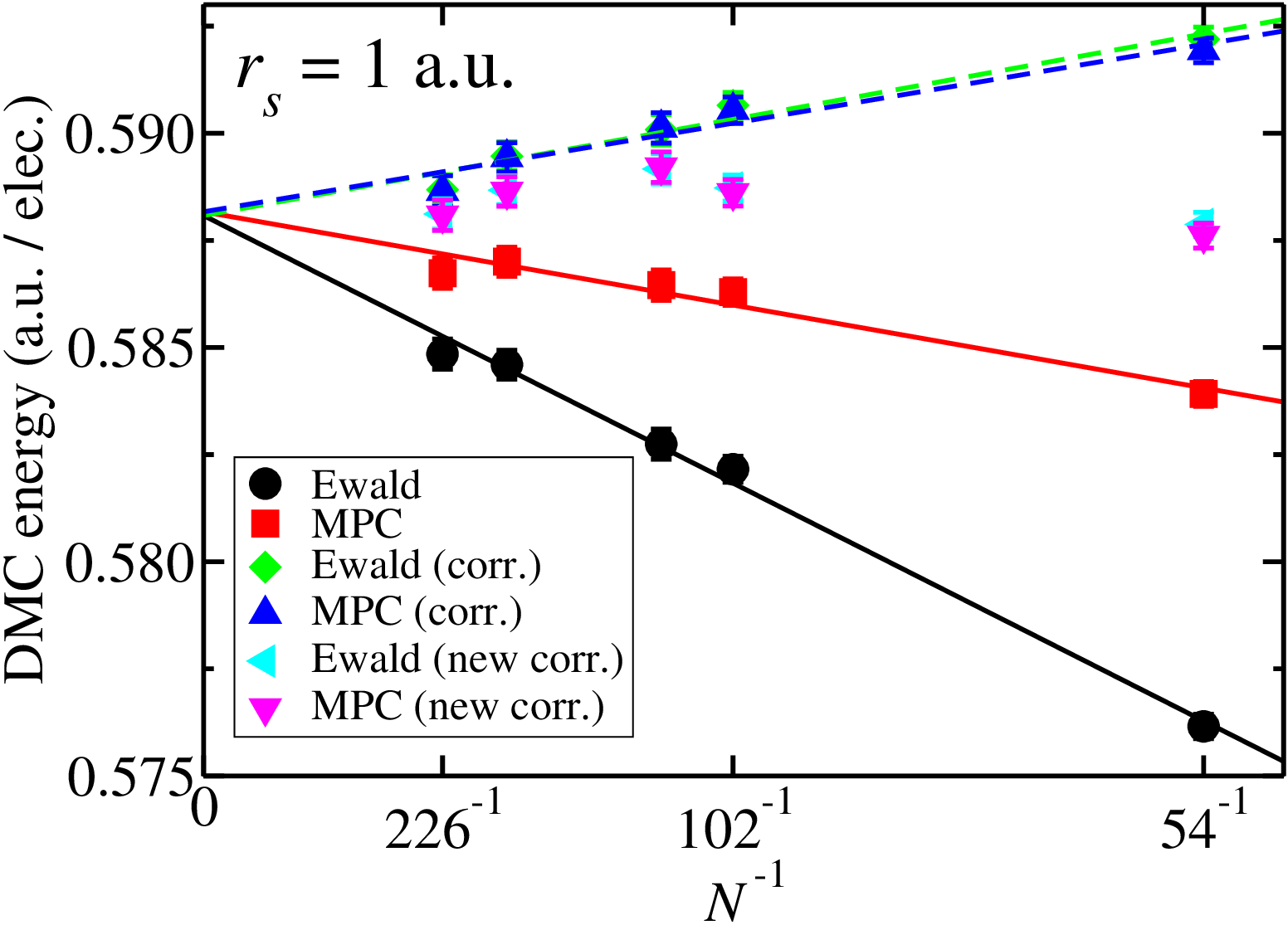} \\[1ex]
\includegraphics[scale=0.3,clip]{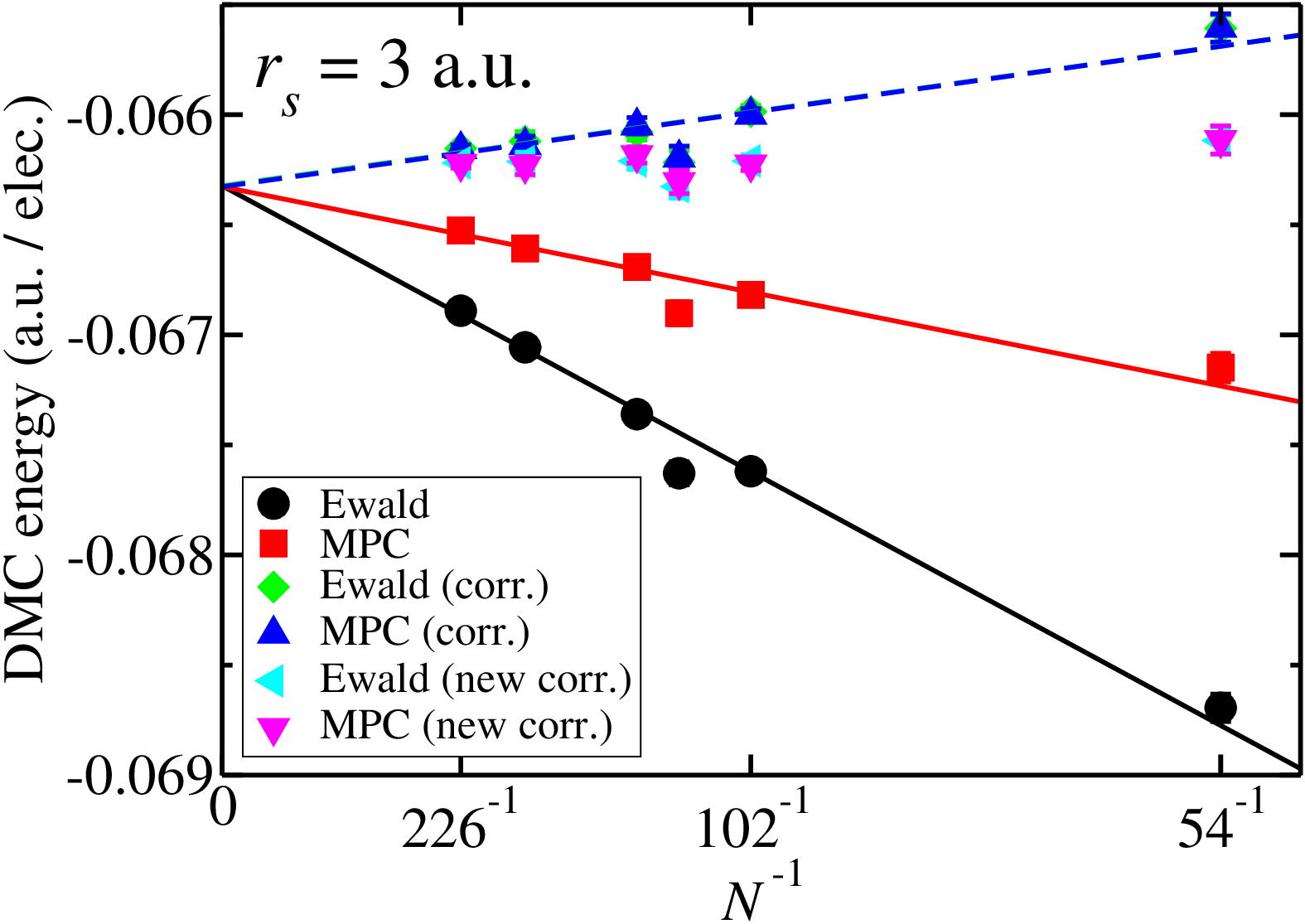} \\[1ex]
\includegraphics[scale=0.3,clip]{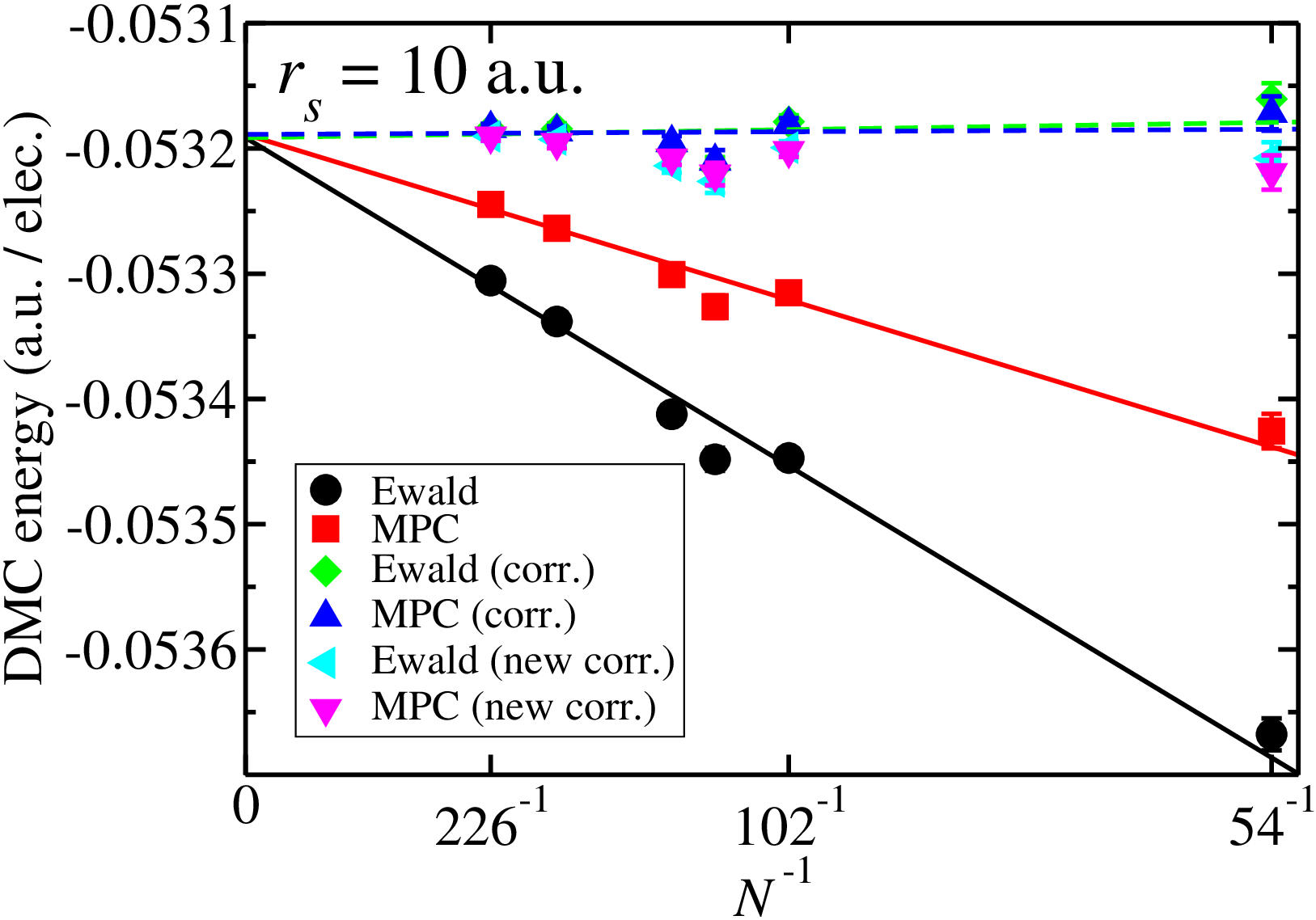}
\caption{(Color online) DMC total energy per electron against the reciprocal
of the number of electrons in 3D paramagnetic HEGs of density parameter
$r_s=1$ a.u.\ (top panel), $r_s=3$ a.u.\ (middle panel), and $r_s=10$ a.u.\
(bottom panel).  The simulation parameters were as described in the caption of
Table \ref{table:heg_ewald_v_mpc}.  The Ewald energy per electron is corrected
by the addition of $(\Delta V_{\rm Ew}+\Delta T_A)/N=\omega_p/(2N)$ [``Ewald
(corr.)''], while the MPC energy is corrected by the addition of $\Delta
T_A/N=\omega_p/(4N)$ [``MPC (corr.)'']. The corrected Ewald and MPC energies
are hard to distinguish because they lie almost on top of each other. The
higher-order KE corrections described in Sec.\ \ref{sec:new_ke_corr} ($\Delta
T_B/N$ plus the single-particle correction) are included in the data sets
labeled ``(new corr.)''.
\label{fig:E_v_N_HEG}}
\end{center}
\end{figure}

The need to include higher-order finite-size corrections to the KE is
demonstrated in Fig.\ \ref{fig:E_v_N_HEG}, which shows the size dependence of
the DMC energy of the 3D HEG\@.  The XC- and KE-corrected Ewald data and the
KE-corrected MPC data are in good agreement with each other, as expected.  At
low density ($r_s=10$ a.u.)\ the corrected data are almost independent of
system size, indicating that the finite-size correction formulas are working
well.  However, at intermediate ($r_s=3$ a.u.)\ and high density ($r_s=1$
a.u.)\ it is clear that the QMC data are overcorrected when only the
leading-order KE correction is applied.  Since the finite-size correction
$\Delta V_{\rm Ew}$ to the interaction energy has been shown to be accurate,
the problem must lie in the KE\@.  It is clearly necessary to go beyond
leading order when correcting the KE at intermediate and high densities.

The Poisson summation formula can be used to demonstrate that higher-order
terms are more important in the KE than the Ewald energy.  If we assume that
the XC hole is well localized within the simulation cell and that $\lim_{{\bf
k}\rightarrow{\bf 0}} k^2 u(k)$ exists, the finite-size correction to the KE
may be obtained from Eq.\ (\ref{eqn:expec_T}) as
\begin{eqnarray}
\Delta T & = & \frac{N}{4} \left( \frac{1}{(2 \pi)^3} \int_{k<\infty} k^2
u({\bf k}) [S({\bf k})-1] \, d{\bf k} - \frac{1}{\Omega} \sum_{{\bf G}_s \neq
{\bf 0}} G_s^2 u({\bf G}_s) [S({\bf G}_s)-1] \right) \\ & = & \frac{N}{4}
\left( - \frac{1}{\Omega} \lim_{{\bf k}\rightarrow{\bf 0}} k^2 u({\bf k}) -
\sum_{{\bf R}_s \neq {\bf 0}} L({\bf R}_s) \right),
\end{eqnarray}
where we have used the Poisson summation formula\cite{footnote:poisson} and
\begin{equation}
L({\bf r})=- \nabla^2 \int_{r^\prime<\infty} u({\bf r}-{\bf r}^\prime)
\rho_{\rm xc}({\bf r}^\prime) \, d{\bf r}^\prime
\end{equation}
is the inverse Fourier transform of $k^2 u({\bf k})[S({\bf k})-1]$.

The leading-order behavior of the two-body Jastrow factor of a HEG at small
$k$ is\cite{gaskell} $u(k)= -4\pi \left(A/k^2+B/k \right)$ within the
RPA\@. Hence, at large $r$, $u(r)=-A/r - 2B/(\pi r^2)$ and so $\nabla^2 u({\bf
r})=-4B/(\pi r^4)$ for $r \neq 0$.  The finite-size correction to the KE is
therefore
\begin{equation}
\Delta T = \frac{N \pi A}{\Omega} - \frac{N B}{\pi} \sum_{{\bf R}_s \neq {\bf
0}} \int_{r<\infty} \frac{\rho_{\rm xc}({\bf r}) \, d{\bf r}}{|{\bf R}_s-{\bf
r}|^4}.
\end{equation}
The first term is the correction of Eq.\ (\ref{eqn:lead_order_ke}), while the
second term gives an additional correction that falls off  slowly as
$N^{-1/3}$. So, even in the case of the HEG, where the next-to-leading-order
correction to the Ewald energy falls off as $N^{-1}$, higher-order corrections
to the KE may be important.

The additional KE correction is due to the discontinuous gradient of $k^2
u({\bf k})$ at ${\bf k}={\bf 0}$.  A similar approach to that developed in
Sec.\ \ref{sec:sol_na} can be used to eliminate the leading-order nonanalytic
contributions to the long-ranged part of $\nabla^2u({\bf r})$.  Define $F({\bf
k}) = k^2 u({\bf k}) [S({\bf k}) - 1]$ and write $F({\bf k})=F_a({\bf
k})+F_b({\bf k})$, where $F_b({\bf k})$ contains the $O(k)$ contribution to
$-k^2u({\bf k})$ [as well as any anisotropic $O(k^{0})$ terms], and is smooth
and long-ranged in ${\bf k}$-space.  Then
\begin{equation}
  \Delta T = \frac{N}{4} \left( \frac{1}{\Omega} \lim_{{\bf k}\rightarrow {\bf
  0}} F_a({\bf k}) + \frac{1}{(2\pi)^3} \int_{k<\infty} F_b({\bf k}) \, d{\bf
  k}-\frac{1}{\Omega} \sum_{{\bf G}_s\neq{\bf 0}} F_b({\bf G}_s) \right) +
  \frac{N}{4}\sum_{{\bf R}_s \neq {\bf 0}} F_a({\bf R}_s).
  \label{eqn:delta_T_full}
\end{equation}

Note that, as shown in Sec.\ \ref{sec:twistav_comp}, the bias due to residual
CE twist-averaged single-particle KE errors also falls off as $N^{-1/3}$.  If
we include higher-order corrections for the neglect of long-ranged
correlations, we should also correct for the residual error in the
twist-averaged energy.

\subsection{Higher-order KE corrections \label{sec:new_ke_corr}}

Gaskell\cite{gaskell} has derived the following expression for the small-$k$
limit of $u({\bf k})$ for the 3D HEG within the RPA:
\begin{eqnarray}
u(k) & = & - \frac{\Omega}{N} \left\{-\frac{1}{2S_0(k)} + \left[
\frac{1}{4S_0^2(k)} + \left( \frac{v_E(k) N}{\Omega \omega_p} \right)^2
\right]^{1/2} \right\} \label{eqn:full_urpa} \\ & \equiv & -4 \pi \left[
\frac{A}{k^2} + \frac{B}{k} \right] + O(k^0),
\label{eqn:approx_urpa}
\end{eqnarray}
where
\begin{equation}
S_0(k) = \sum_{\sigma} \frac{N_{\sigma}}{N} \left[  \frac{3k}{4k_{{\rm
F}\sigma}} - \frac{k^3}{16k_{{\rm F}\sigma}^3} \right]
\end{equation}
is the HF SF, $k_{{\rm F} \sigma}=\left( 6 \pi^2 N_{\sigma} / \Omega
\right)^{1/3}$ is the Fermi wave vector for particles of spin $\sigma$,
$N_{\sigma}$ is the number of particles of spin $\sigma$, and
\begin{eqnarray}
A & = & \frac{1}{\omega_p} = \sqrt{\frac{r_s^3}{3}} \\ B & = & -
\frac{2r_s^2}{3} \left( \frac{2 \pi}{3} \right)^{1/3} \left[
(1+\zeta)^{2/3}+(1-\zeta)^{2/3} \right]^{-1},
\label{eqn:A_heg}
\end{eqnarray} where
$\zeta=(N_\uparrow-N_\downarrow)/N$ is the spin polarization.

Let $F_b(k)=4\pi B k \exp(-\alpha k^2)$.  This satisfies the requirements for
$F_b$ given in Sec.\ \ref{sec:need_for_ke_corr}, provided $\alpha$ is small.
Then, by Eq.\ (\ref{eqn:delta_T_full}) in the limit $\alpha\rightarrow 0$,
\begin{eqnarray}
\Delta T & = & \frac{N}{4} \left( \frac{4\pi A}{\Omega} + \frac{B}{\pi
  \alpha^2} - \frac{4 \pi B}{\Omega} \sum_{{\bf G}_s \neq {\bf 0}} G_s \exp
  \left(-\alpha G_s^2\right) \right) + O(N^{-2/3}) \nonumber \\ & = & \frac{N
  \pi A}{\Omega} + \frac{C_{\rm 3D} NB}{\Omega^{4/3}} + O(N^{-2/3})
  \label{eqn:gen:_KE_corr} \\ & = & \frac{\omega_p}{4} - \frac{C_{\rm
  3D}}{2\pi r_s^2 (2N)^{1/3}} \left[ (1+\zeta)^{2/3}+(1-\zeta)^{2/3}
  \right]^{-1} + O(N^{-2/3}) \equiv \Delta T_A + \Delta T_B + O(N^{-2/3}),
\end{eqnarray} where $C_{\rm 3D}=5.083$, $5.264$, and $5.086$ for FCC, SC, and
BCC simulation cells, respectively.  The $O(N^{-2/3})$ error arises from the
$O(r^{-3})$ term in $u(r)$ at large $r$.

The relative importance of the corrections for typical system sizes at three
different densities is shown in Table \ref{table:size_new_corrs}. The residual
CE twist-averaged single-particle KE error is generally greater than $\Delta
T_B$.  This error can be estimated within HF theory.\cite{footnote_effmass}
For real systems, the ``infinite-system'' HF energy would have to be evaluated
in a large, finite calculation.  The effect of adding higher-order corrections
(including the correction for the residual single-particle error) to the
energy of a 3D HEG is demonstrated in Fig.\ \ref{fig:E_v_N_HEG}.  The
finite-size behavior of the QMC data is clearly greatly improved at $r_s=1$
and 3 a.u.

\begin{table}
\begin{tabular}{ccr@{.}lr@{.}lr@{.}l}
\hline \hline

& & \multicolumn{6}{c}{KE correction (a.u.\ / elec.)} \\

\raisebox{1ex}[0pt]{$r_s$ (a.u.)} & \raisebox{1ex}[0pt]{$N$} &
\multicolumn{2}{c}{SP corr.} & \multicolumn{2}{c}{$\Delta T_A/N$} &
\multicolumn{2}{c}{$\Delta T_B/N$} \\

\hline

1  & ~54 & $-0$&$002\,8$   & ~$0$&$008\,0$  & $-0$&$001\,6$ \\

1  & 130 & $-0$&$000\,65$  & $0$&$003\,33$  & $-0$&$000\,48$ \\

3  & ~54 & $-0$&$000\,31$  & $0$&$001\,54$  & $-0$&$000\,17$ \\

3  & 130 & $-0$&$000\,072$ & $0$&$000\,641$ & $-0$&$000\,054$ \\

10 & ~54 & $-0$&$000\,027$ & $0$&$000\,254$ & $-0$&$000\,015$ \\

10 & 130 & $-0$&$000\,006$ & $0$&$000\,105$ & $-0$&$000\,005$ \\

\hline \hline
\end{tabular}
\caption{Magnitude of different components of the finite-size correction to
the KE per electron of a 3D paramagnetic HEG at different density parameters
$r_s$ and system sizes $N$ in an FCC cell. The correction for residual
single-particle errors after twist averaging in the CE (``SP corr.'')\ is
estimated as the difference between the infinite-system HF KE and the
twist-averaged HF KE for the finite system.
\label{table:size_new_corrs}}
\end{table}

For real systems we do not usually have an analytic result for the small-$k$
behavior of $u({\bf k})$.  However, we have flexible forms of $u({\bf r})$
that can be optimized within QMC\@.  By fitting a suitable functional form to
the QMC-optimized $u$, we can extrapolate to the ${\bf k}={\bf 0}$ limit.  We
suggest that Eq.\ (\ref{eqn:approx_urpa}) be fitted to the QMC $u(G_s)$ at the
first two stars of nonzero ${\bf G}_s$ vectors, and that Eq.\
(\ref{eqn:gen:_KE_corr}) should then be used to evaluate the KE correction.

\subsection{Low-${\bf k}$ behavior of the Fourier-transformed two-body
Jastrow factor}

\begin{figure}
\begin{center}
\includegraphics[scale=0.3,clip]{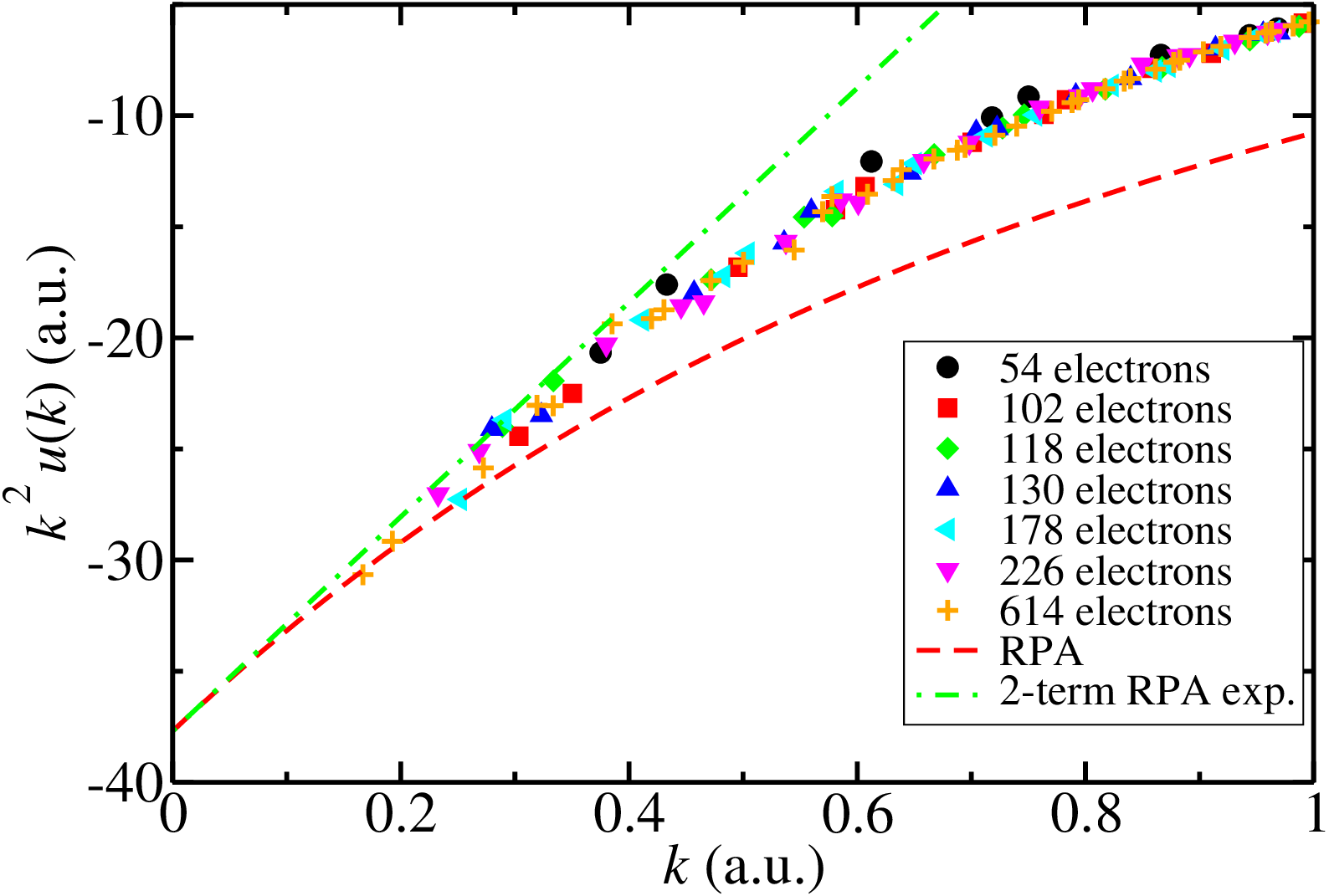}
\caption{(Color online) Fourier transform of the fully optimized two-body
  Jastrow factor for a paramagnetic 3D HEG at $r_s=3$ a.u.\ and different
  system sizes. For comparison, the RPA expression of Eq.\
  (\ref{eqn:full_urpa}) and the two-term RPA expansion [Eq.\
  (\ref{eqn:approx_urpa})] are shown. \label{fig:FT_of_u_rs3}}
\end{center}
\end{figure}

\begin{figure}
\begin{center}
\includegraphics[scale=0.3,clip]{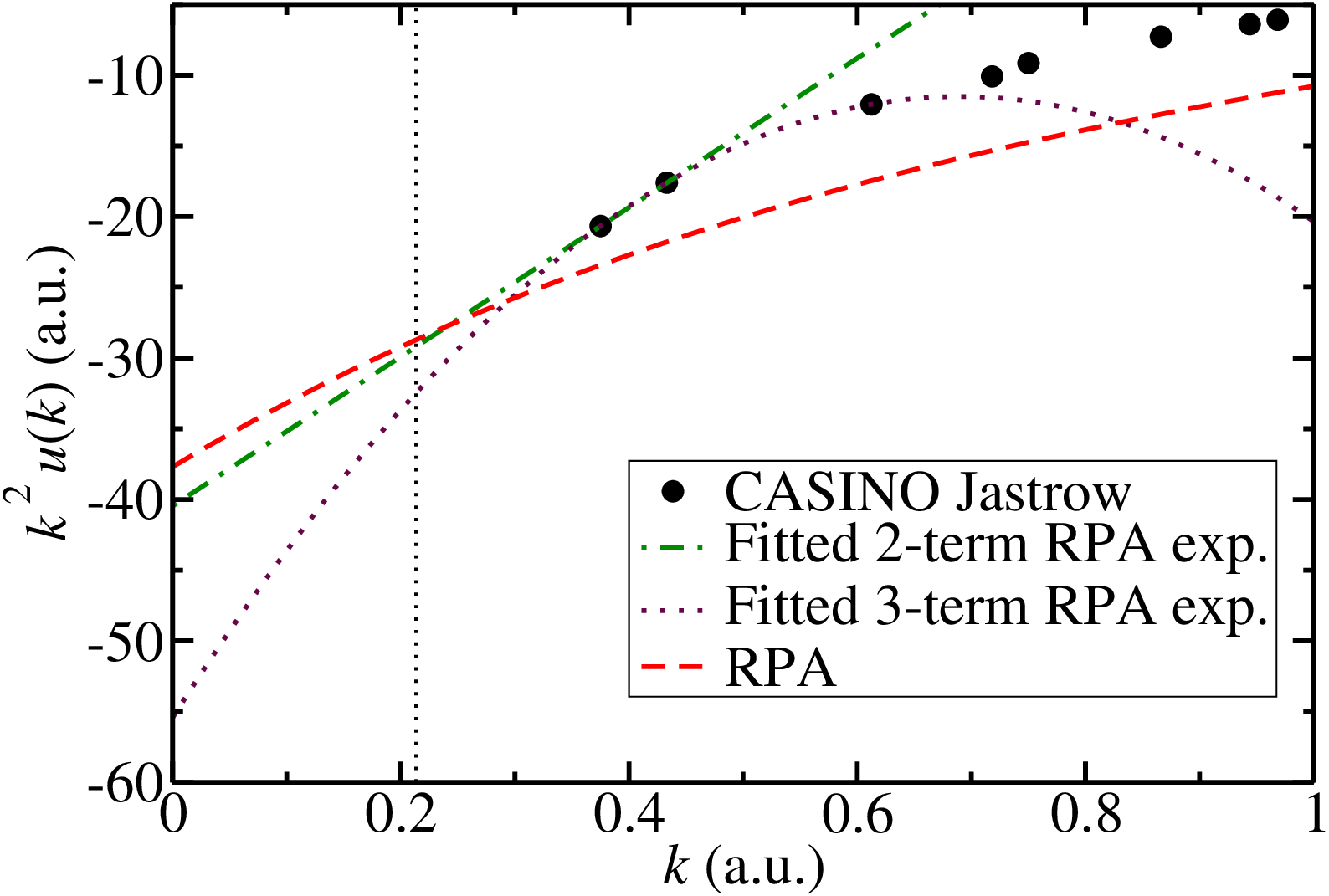} \\[1ex]
\includegraphics[scale=0.3,clip]{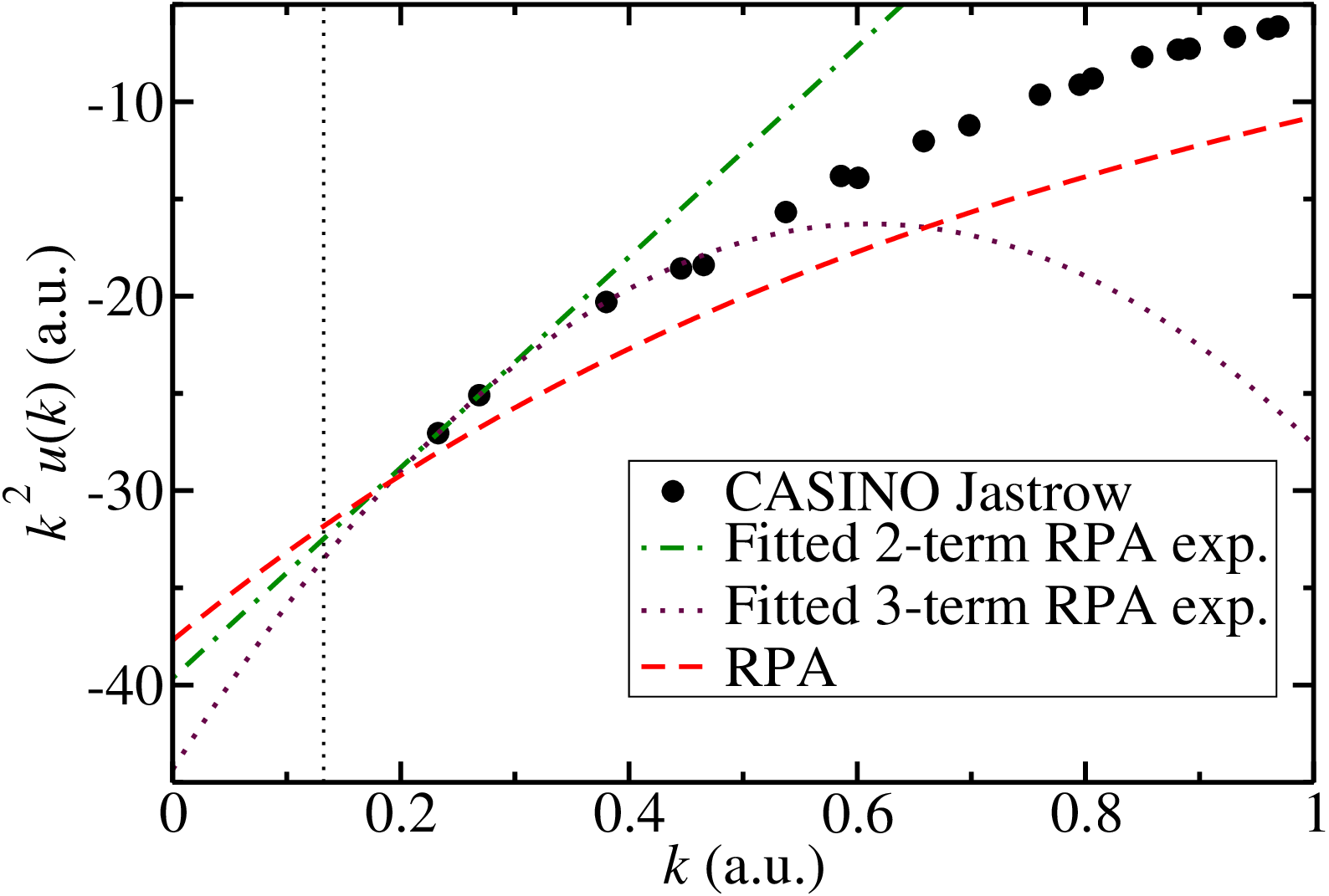}
\caption{(Color online) Two-body Jastrow factor for a paramagnetic 3D HEG at
  $r_s=3$ a.u.\ with 54 electrons (top panel) and 226 electrons (bottom
  panel).  The Jastrow factors are as in Fig.\ \ref{fig:FT_of_u_rs3}.  The RPA
  expression of Eq.\ (\ref{eqn:full_urpa}) is plotted, as are fits of the two-
  and three-term RPA expansions [Eq.\ (\ref{eqn:approx_urpa}) and Eq.\
  (\ref{eqn:approx_urpa}) with an extra term $-4\pi C$]. The fits were made to
  the QMC-optimized $u$ at the first two and first three stars of ${\bf G}_s$
  vectors, respectively. The dotted line indicates the radius of the sphere
  whose volume is $(2\pi)^3/\Omega$. \label{fig:FT_of_u_HEG_rs3_N54_N226}}
\end{center}
\end{figure}

The Fourier transform of the two-body Jastrow factor of a 3D paramagnetic HEG
at $r_s=3$ a.u.\ is shown in Fig.\ \ref{fig:FT_of_u_rs3}. The Jastrow factor
consisted of polynomial and plane-wave expansions in electron-electron
separation,\cite{ndd_jastrow} which were optimized by variance
minimization\cite{umrigar_1988a,ndd_newopt} followed by energy
minimization.\cite{umrigar_emin} As expected, the form of $u({\bf k})$ is
largely independent of the number of electrons, and the small-$k$ behavior is
well-described both by the RPA of Eq.\ (\ref{eqn:full_urpa}) and by the first
two terms of the power-series expansion, Eq.\ (\ref{eqn:approx_urpa}).  The
RPA expression for the two-body Jastrow factor does not satisfy the Kato cusp
conditions\cite{kato_pack} and hence becomes unreliable at large $k$. The
small-$k$ behavior of $u({\bf k})$ for 54- and 226-electron HEGs is shown in
Fig.\ \ref{fig:FT_of_u_HEG_rs3_N54_N226}.  It can be seen that the fit of the
two-term RPA expansion to the QMC-optimized Jastrow factor is a fairly good
approximation to the analytic RPA form within a sphere of volume
$(2\pi)^3/\Omega$, but that the fitted three-term RPA expansion is badly
behaved, because one is simply fitting to the noise in the $u({\bf G}_s)$
data.  This is reflected in the corresponding results for the KE correction
shown in Table \ref{table:fitting_forms}. The corrections obtained with the
fitted two-term expansion are close to the analytic KE correction
(leading-order and next-to-leading order terms).  The leading-order correction
can be thought of as being calculated on the assumption that $k^2u(k)$ is
constant over the integration regions shown in Fig.\
\ref{fig:FT_of_u_HEG_rs3_N54_N226}, which is clearly inappropriate, and leads
to the overcorrection for $N=54$ electrons.  The fitted three-term RPA
expansion also gives an overcorrection. For HEGs, the analytic results given
in Sec.\ \ref{sec:new_ke_corr} should of course be used.

\begin{table}
\begin{center}
\begin{tabular}{lcr@{}lr@{}l}
\hline \hline

Method & $N$ & \multicolumn{2}{c}{$A$ (a.u.)} & \multicolumn{2}{c}{$\Delta T$
  (a.u.)} \\

\hline

Analytic RPA         & $54$  & ~~$3$&    & ~~$0$&$.073\,9$ \\

Analytic 1-term exp. & Any   & $3$&      & $0$&$.083\,2$ \\

Fitted 2-term exp.   & $54$  & $3$&$.22$ & $0$&$.079\,1$ \\

Fitted 3-term exp.   & $54$  & $4$&$.41$ & $0$&$.097\,6$ \\

Analytic RPA         & $226$ & $3$&      & $0$&$.077\,5$ \\

Fitted 2-term exp.   & $226$ & $3$&$.15$ & $0$&$.081\,0$ \\

Fitted 3-term exp.   & $226$ & $3$&$.53$ & $0$&$.086\,9$ \\

\hline \hline
\end{tabular}
\caption{Finite-size correction to the KE of a paramagnetic HEG of density
parameter $r_s=3$ a.u., calculated using different two-body Jastrow factors.
The value of $A=3$ corresponds to $1/\omega_p$; see Eq.\ (\ref{eqn:A_heg}). We
compare analytic results with those obtained by fitting to the QMC-optimized
two-body Jastrow factors shown in Fig.\ \ref{fig:FT_of_u_HEG_rs3_N54_N226}.
The ``Analytic RPA'' form is that of Eq.\ (\ref{eqn:full_urpa}), the ``3-term
exp.''\ is that of Eq.\ (\ref{eqn:approx_urpa}) with an extra term $-4\pi C$,
the ``2-term exp.''\ is Eq.\ (\ref{eqn:approx_urpa}), and the ``1-term exp.''\
is Eq.\ (\ref{eqn:approx_urpa}) with $B=0$.
\label{table:fitting_forms}}
\end{center}
\end{table}

\section{Finite-size corrections in 2D systems \label{sec:fs_2D}}

\subsection{XC energy in 2D}

Consider a 2D-periodic system with simulation-cell area $P$.  For a
sufficiently symmetric system, $S({\bf k})=\gamma k^{3/2}+O(k^2)$.\cite{pines}
Hence $\lim_{{\bf k}\rightarrow{\bf 0}} v_E(k) {S}({\bf k}) = 2 \pi \lim_{{\bf
k}\rightarrow{\bf 0}} {S}(k)/k=0$.  So the 2D analog of Eq.\
(\ref{eqn:general_xc_corr}) is
\begin{equation}
\Delta V_{\rm Ew} = -\frac{N}{2} \sum_{{\bf R}_s \neq {\bf 0}} \phi_{\rm
loc}({\bf R}_s).
\label{eqn:lr_xc_2d}
\end{equation}
In a 2D HEG the nonoscillatory XC hole is relatively long-ranged due to the
reduced screening, decaying as $\rho_{\rm xc}(r)=-\tilde{\Lambda} r^{-7/2}$,
where $\tilde{\Lambda}$ is a constant.\cite{gorigiorgi_2D} Hence the XC charge
outside radius $r$ is $-4 \pi \tilde{\Lambda}/(3r^{3/2})$ and the leading
(monopolar) contribution to $\phi_{\rm loc}({\bf r})$ is proportional to
$r^{-5/2}$ at large $r$. [The dipole moment of the electron and its XC hole is
zero, while the quadrupole\cite{footnote:2D_quadrupole} contribution to
$\phi_{\rm loc}({\bf r})$ is proportional to $r^{-3}$.] Hence
\begin{equation}
\Delta V_{\rm Ew}  \propto  - \frac{N}{2} \sum_{{\bf R}_s \neq 0} R_s^{-5/2} =
O(N^{-1/4}) ,
\end{equation}
since the length of every simulation-cell lattice vector ${\bf R}_s$ appearing
in the summation is proportional to $\sqrt{N}$. Unlike the 3D case, therefore,
$\Delta V_{\rm Ew} \rightarrow 0$ as $N \rightarrow \infty$.  This conclusion
was also reached, using a different approach, by Wood \textit{et
al.}\cite{wood}

To obtain the leading-order correction to the XC energy, we use the method of
Sec.\ \ref{sec:sol_na}.  Let $S_b(k)=\gamma k^{3/2} \exp(-\alpha k^2)$.  Then,
by the 2D analog of Eq.\ (\ref{eqn:isolation}),
\begin{eqnarray} \Delta V_{\rm Ew} & = & \frac{N}{2} \left( 0+\frac{1}{(2\pi)^2} \int_0^{\infty}
  v_E(k) S_b(k) \times 2\pi k \, dk - \frac{1}{P} \sum_{{\bf G}_s \neq {\bf
  0}} v_E(G_s) S_b(G_s) \right) + O(N^{-1/2}) \nonumber \\ & = & \frac{N}{2}
  \left( \frac{\Gamma(5/4) \gamma}{2 \alpha^{5/4}} - \frac{2\pi \gamma}{P}
  \sum_{{\bf G}_s\neq {\bf 0}} \sqrt{G_s} \exp \left(-\alpha G_s^2 \right)
  \right) + O(N^{-1/2}) \nonumber \\ & = & \frac{C_{\rm 2D}  N\gamma}{P^{5/4}}
  + O(N^{-1/2}),
\end{eqnarray}
where $C_{\rm 2D}=3.9852$ and 3.9590 for square and hexagonal cells,
respectively, and the $\alpha \rightarrow 0$ limit was taken in the final
step.  The $O(N^{-1/2})$ error is due to the quadrupole moment of $S_a({\bf
r})=S({\bf r})-S_b({\bf r})$.  For a 2D HEG,\cite{gorigiorgi_2D}
$\gamma=2^{-3/4}r_s^{-1/2}$. Hence
\begin{equation}
\Delta V_{\rm Ew}=\frac{C_{\rm 2D}}{2\pi r_s^3} \left( \frac{2}{\pi N}
  \right)^{1/4} + O(N^{-1/2}). \label{eqn:2d_xc_corr}
\end{equation}
This correction falls off very rapidly with $r_s$.

\subsection{KE in 2D}

For a symmetric 2D-periodic system,\cite{tanatar_1989} $u(k)=-ak^{-3/2}
+O(k^{-1})$ and $S(k) = \gamma k^{3/2} + O(k^2)$.  Hence, proceeding as in
Sec.\ \ref{sec:need_for_ke_corr}, we have $F(k) = k^2 u(k) [S(k) - 1] = a
k^{1/2} + O(k)$. Let $F_b(k)=ak^{1/2}\exp(-\alpha k^2)$.  Then, by the 2D
analog of Eq.\ (\ref{eqn:delta_T_full}),
\begin{equation}
\Delta T = \frac{N}{4} \left( 0 + \frac{\Gamma(5/4)a}{4 \pi \alpha^{5/4}} -
  \frac{a}{P} \sum_{{\bf G}_s \neq {\bf 0}} \sqrt{G_s} \exp \left(-\alpha
  G_s^2 \right) \right) + O(N^{-1/2}) = \frac{C_{\rm 2D}Na}{4\pi P^{5/4}} +
  O(N^{-1/2}),
\label{eqn:ke_corr2D_simple}
\end{equation}
where the $\alpha\rightarrow 0$ limit was taken in the final step.

For a 2D HEG the HF SF is\cite{giuliani}
\begin{equation}
S_0(k) = \sum_\sigma \frac{2N_\sigma}{\pi N} \left[ \sin^{-1}\left(
  \frac{k}{2k_{{\rm F} \sigma}} \right) + \frac{k}{2k_{{\rm F} \sigma}}
  \sqrt{1-\left( \frac{k}{2k_{{\rm F} \sigma}} \right)^2} \right],
\end{equation}
where the Fermi wave vector for electrons of spin $\sigma$ is $k_{{\rm F}
\sigma}=\sqrt{4\pi N_\sigma / P}$.  The small-$k$ limit of the two-body
Jastrow factor within the RPA is\cite{tanatar_1989}
\begin{equation}
u(k)=- \frac{P}{N} \left\{ \frac{-1}{2S_0(k)} + \left[ \frac{1}{[2S_0(k)]^2} +
\frac{Nv_E(k)}{P k^2} \right]^{1/2} \right\} = -\frac{\sqrt{2} \pi
r_s}{k^{3/2}} + O(k^{-1}).
\end{equation}
Hence
\begin{equation}
\Delta T = \frac{C_{\rm 2D}}{\pi(\pi N)^{1/4} (2 r_s)^{3/2}} + O(N^{-1/2}).
\label{eqn:ke_corr2D_heg}
\end{equation}

For real systems, we suggest that the Fourier transform of the two-body
Jastrow factor be fitted to $u(k)=-a/k^{3/2}-b/k$ using the first two nonzero
stars of simulation-cell ${\bf G}_s$ vectors. Equation
(\ref{eqn:ke_corr2D_simple}) should then be used to calculate the KE
correction.

\subsection{Effectiveness of 2D KE correction}

\begin{figure}
\begin{center}
\includegraphics[scale=0.3,clip]{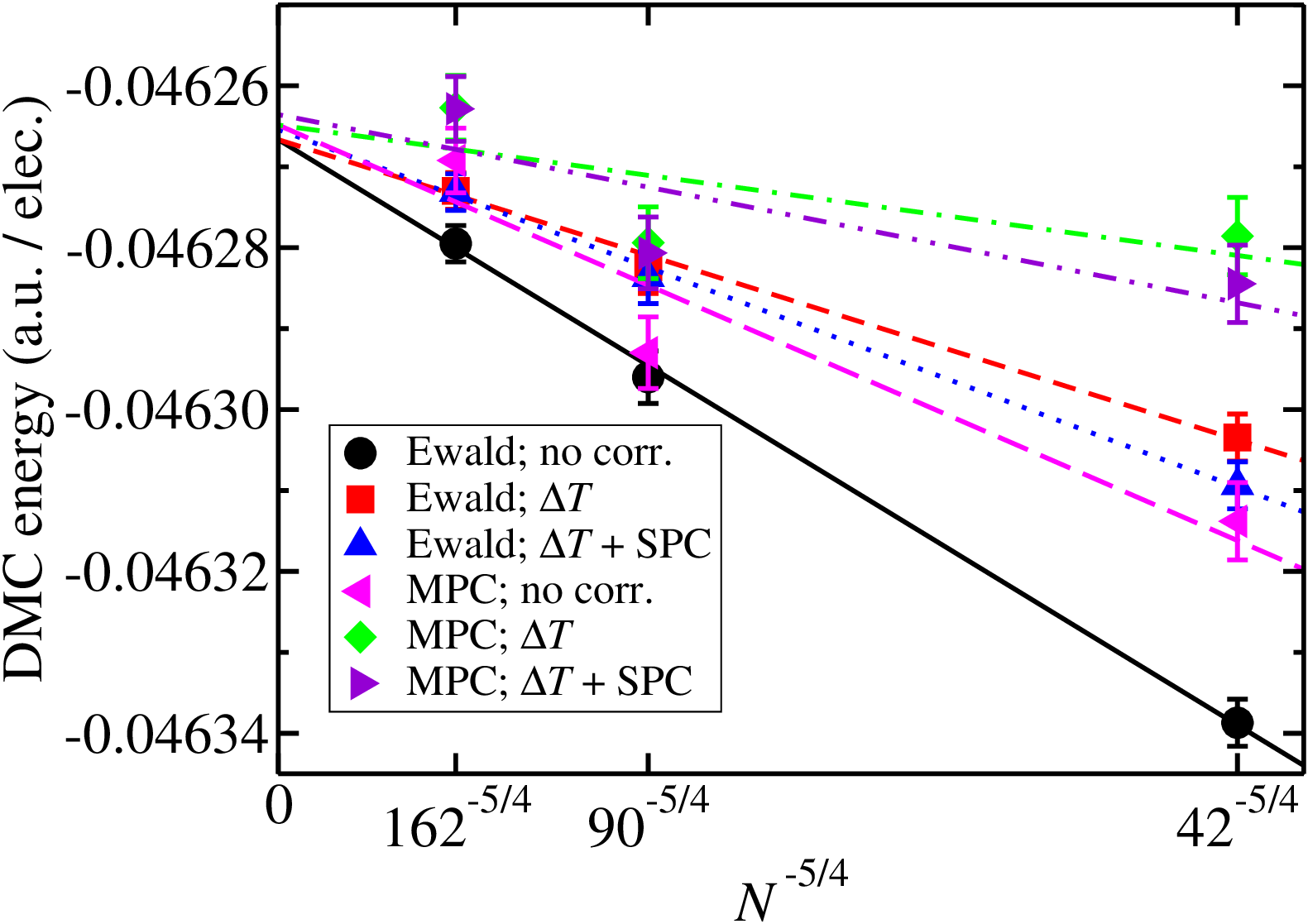}
\caption{(Color online) Twist-averaged DMC energy per electron against system
size for a 2D paramagnetic HEG of density parameter $r_s=20$\ a.u.  The trial
wave function was of Slater-Jastrow-backflow form,\cite{backflow} the target
population was 1536 configurations, and the DMC energies have been
extrapolated to zero time step. In each case the Ewald interaction was used in
the DMC branching factor.  ``Ewald'' and ``MPC'' indicate the interaction used
to calculate the local energies.  $\Delta T$ is given in Eq.\
(\ref{eqn:ke_corr2D_heg}), respectively, and ``SPC'' denotes the
single-particle correction to the KE (the difference of the infinite-system
and CE twist-averaged finite-system HF KE's). \label{fig:para_rs20_E_v_N}}
\end{center}
\end{figure}

We illustrate the effectiveness of the KE corrections in a 2D HEG at low
density in Fig.\ \ref{fig:para_rs20_E_v_N}.  The XC correction [Eq.\
(\ref{eqn:2d_xc_corr})] is negligibly small at this density. However it is
clear that applying finite-size corrections to the KE alone is not sufficient
to obtain accurate results. The MPC interaction gives significantly smaller
finite-size errors than the Ewald interaction; nevertheless it is clear that
extrapolation is necessary.

\section{Formulas for finite-size extrapolation
  \label{sec:extrapolation_expressions}}

\subsection{Finite-size extrapolation}

In nearly all QMC studies of condensed matter to date it has been necessary to
extrapolate energy data to infinite system size by means of an assumed
relationship between energy and particle number.  These formulas contain free
parameters, including the infinite-system energy, which are determined by a
fit to the QMC data.  Despite the existence of sophisticated methods for
treating finite-size errors, it is likely that some form of extrapolation will
continue to be necessary for accurate work. In this section we analyze the
performance of fitting formulas that have been proposed in the literature and
consider how best to extrapolate QMC energies to infinite system size.

Throughout this section we denote the QMC energy per electron of an
$N$-electron system by $e(N)$ and we denote the HF energy, KE, and interaction
energy per electron by $e_{\rm HF}(N)$, $t_{\rm HF}(N)$, and $v_{\rm HF}(N)$,
respectively.  We assume that the same ${\bf k}_s$ is used in both the QMC and
HF calculations (or that twist averaging is applied in both cases).

\subsection{Finite-size extrapolation formulas for the HEG}

The exact size-dependence of the HF energy of the fluid phase of the HEG is
\begin{equation}
e_{\rm HF}(\infty) = e_{\rm HF}(N) + \Delta t_{\rm HF}(N) + \Delta v_{\rm
HF}(N),
\label{eqn:extrap_formula_hf}
\end{equation}
where $\Delta t_{\rm HF}(N)=t_{\rm HF}(\infty)-t_{\rm HF}(N)$ and $\Delta
v_{\rm HF}(N)=v_{\rm HF}(\infty)-v_{\rm HF}(N)$.  The forms of $\Delta t_{\rm
HF}(N)$ and $\Delta v_{\rm HF}(N)$ for a 3D paramagnetic HEG can be seen in
Fig.\ \ref{fig:twist_convergence}.  Both are oscillatory functions of $N$ due
to single-particle finite-size errors.  The fluctuations in the exchange
energy and the KE are strongly correlated, although those in the KE are
larger.  For further discussion of the single-particle finite-size errors in
HF theory, see Sec.\ \ref{subsec:twist-averaging} and Ref.\
\onlinecite{lin_twist_2001}.  There is also a systematic error in the HF
exchange energy, caused by the compression of the exchange hole, as discussed
in Appendix \ref{sec:hf_fs_errors}.  For a Wigner crystal,
Ceperley\cite{ceperley_1978} suggested the fitting form
\begin{equation}
e(\infty)=e(N)+\frac{c}{r_s^{3/2}N^{3/d}},
\label{eqn:cep_extrap_formula_wigner}
\end{equation}
where $d$ is the dimensionality and $c$ is roughly independent of $r_s$.  This
is consistent with the form of the 3D XC correction [Eq.\ (\ref{eq:finsize1})]
and the leading-order correction to the KE [Eq.\ (\ref{eqn:lead_order_ke})].
For an interacting Fermi fluid, Ceperley\cite{ceperley_1978} suggested that
the HF extrapolation is appropriate at small $r_s$, while the Wigner-crystal
extrapolation is more reasonable at large $r_s$.  He therefore proposed using
an interpolation of Eqs.\ (\ref{eqn:extrap_formula_hf}) and
(\ref{eqn:cep_extrap_formula_wigner}),
\begin{equation}
e(\infty) = e(N) + \Delta t_{\rm HF}(N) + \left(\frac{1}{\Delta v_{\rm HF}(N)}
+ \frac{N^{3/d} r_s^{3/2}}{c} \right)^{-1}.
\label{eqn:extrap_formula_ceperley_1978}
\end{equation}

For their study of the 3D HEG, Ceperley and Alder\cite{ceperley_1980} used the
two-parameter form
\begin{equation}
e(\infty) = e(N) + a \Delta t_{\rm HF}(N) + \frac{c}{N^{3/d}},
\label{eqn:extrap_formula_tanatar_1989}
\end{equation}
where $a$ and $c$ are fitting parameters that vary with density.  The
parameter $a$ may be thought of as the ratio of the actual electron mass to
the effective mass within Fermi liquid theory.  One therefore expects $a
\approx 1$ in weakly correlated systems. Alternatively one can estimate $a =
\Delta t(N)/\Delta t_{\rm HF}(N)$.  The parameter $c$ accounts for the Coulomb
finite-size effects in the XC energy and the neglect of long-ranged
correlations in the KE\@. This form has also been used for the 2D
HEG,\cite{tanatar_1989} although our analysis (see Sec.\ \ref{sec:fs_2D})
suggests that a term of the form $cN^{-5/4}$ would be more appropriate than
$cN^{-3/2}$. In their studies of the 3D HEG, Ortiz \textit{et al.}\cite{ortiz}
tested both Eqs.\ (\ref{eqn:extrap_formula_tanatar_1989}) and
(\ref{eqn:extrap_formula_ceperley_1978}).  They found that the two formulas
give very similar results, but that $c$ in Eq.\
(\ref{eqn:extrap_formula_ceperley_1978}) was a strong function of $r_s$.

Unlike the HF energy, the DFT energy does not suffer from long-ranged
finite-size effects.  Finite-size errors in DFT are entirely due to ${\bf
k}$-point sampling errors, i.e., single-particle finite-size effects.  QMC
energy data for real systems can therefore be extrapolated to infinite system
size as
\begin{equation}
e(\infty) = e(N) + a \Delta e_{\rm DFT}(N) + \frac{c}{N^\gamma},
\label{eqn:extrap_dft}
\end{equation}
where $\gamma=1$ in 3D and $\gamma=5/4$ in 2D, and $\Delta e_{\rm DFT}(N)$ is
the difference of the DFT energy per electron in the limit of fine ${\bf
k}$-point sampling and the DFT energy per electron for the set of ${\bf k}$
vectors used in the QMC calculation.

\subsection{Comparison of extrapolation formulas \label{subsec:new_extrap}}

Consider the extrapolation formula
\begin{equation}
e(\infty) = e(N) + a \Delta t_{\rm HF}(N) + b \Delta v_{\rm HF}(N) +
\frac{c}{N^\gamma}
\label{eqn:heg_fs_fit_formula}
\end{equation}
for a 3D system, where $a$, $b$, $c$, and $\gamma$ are parameters to be
determined by fitting, which are allowed to vary with density.  Imposing the
constraint $b=0$ and $\gamma=1$ gives Eq.\
(\ref{eqn:extrap_formula_tanatar_1989}).  The results of fitting Eqs.\
(\ref{eqn:heg_fs_fit_formula}) and (\ref{eqn:extrap_formula_ceperley_1978}) to
DMC data for paramagnetic Fermi fluids at $r_s=1$, $3$, and $8$ a.u.\ are
shown in Table \ref{table_DMC_fs_fits}.

The extrapolated energies can be compared with the infinite-system limit of
the Slater-Jastrow DMC energies obtained using twist averaging at $r_s=1$ and
$3$ a.u., as shown in Fig.\ \ref{fig:E_v_N_HEG} and quoted in the caption to
Table \ref{table_DMC_fs_fits}.  In each case the optimal value of $b$ is
approximately 0, and the $\chi^2$ value does not increase greatly when $b=0$
is imposed.  Setting $a=b$ (i.e., using the HF total energy to extrapolate
away single-particle finite-size errors) gives a very poor fit to the data and
introduces significant bias into the extrapolated energy.  Both of these
effects are caused by the slowly decaying systematic error in $v_{\rm HF}(N)$
due to the long-ranged exchange hole; this error does not have a counterpart
in the QMC data to which the formula is fitted. At high densities the fit can
be improved considerably by allowing $\gamma$ to vary; however the
extrapolated energies are then biased.  It is preferable to impose the known
behavior $\gamma=1$.  Setting the effective mass $a$ equal to $1$, which is
also implicit in Eq.\ (\ref{eqn:extrap_formula_ceperley_1978}), greatly
increases the $\chi^2$ value of the fit, but does not significantly bias the
extrapolated energy, because it simply reduces the amplitude of the
oscillations in the fitted energy.  Using Eq.\
(\ref{eqn:extrap_formula_ceperley_1978}) or Eq.\
(\ref{eqn:heg_fs_fit_formula}) with $a=1$ is unreliable with small numbers of
data points, however.  Furthermore, Eq.\
(\ref{eqn:extrap_formula_ceperley_1978}) is likely to be poor at low density
because of the inclusion of $\Delta v_{\rm HF}(N)$.  Note that where the fits
are good the effective mass ratios $a$ are in good agreement with one another,
and they increase with $r_s$.

In summary, if single-particle finite-size errors are to be removed by
extrapolation using Eq.\ (\ref{eqn:heg_fs_fit_formula}) then only the HF KE
should be used in the extrapolation formula (i.e., $b$ should be 0), and some
attempt should be made to compute the effective mass $a$.  In 3D the exponent
$\gamma$ should be $1$, while in 2D it should be $5/4$.  However, it is
clearly preferable to remove single-particle finite-size effects by twist
averaging, if possible.

\begin{table}
\begin{center}
\begin{tabular}{ccr@{}lr@{}lr@{}lr@{}lr@{}lr@{}l}
\hline \hline

$r_s$ (a.u.) & Constr. & \multicolumn{2}{c}{$a$} & \multicolumn{2}{c}{$b$} &
\multicolumn{2}{c}{$c$} & \multicolumn{2}{c}{$\gamma$} &
\multicolumn{2}{c}{$e(\infty)$ (a.u.\ / elec.)}  &
\multicolumn{2}{c}{$\chi^2$} \\

\hline

$1$ & None & $1$&$.062$ & $0$&$.132$ & $0$&$.329$ & ~~$0$&$.892$ &
~~~~$0$&$.589\,73$ & $8$&$.8$ \\

$1$ & $b=0$ & $1$&$.087$ & $0$& & $0$&$.355$ & $0$&$.833$ & $0$&$.589\,57$ &
$23$&$.1$ \\

$1$ & $b=0$, $\gamma=1$ & $1$&$.084$ & $0$& & $0$&$.532$ & $1$& &
$0$&$.587\,38$ & $3\,290$& \\

$1$ & $a=b$ & $0$&$.924$ & $0$&$.924$ & $-0$&$.050$ & $0$&$.194$ &
$0$&$.578\,39$ & $676$& \\

$1$ & $a=\gamma=1$, $b=0$ & $1$& & $0$& & $1$& & $-0$&$.388$ & $0$&$.584\,96$
& $30\,300$& \\

$1$ & Eq.\ (\ref{eqn:extrap_formula_ceperley_1978}) &
\multicolumn{2}{c}{$\cdots$} & \multicolumn{2}{c}{$\cdots$} & $1$&$.060$ &
\multicolumn{2}{c}{$\cdots$} & $0$&$.587\,86$ & $18\,900$& \\

$3$ & None & $1$&$.107$ & $0$&$.067$ & $0$&$.147$ & $1$&$.040$ &
$-0$&$.066\,10$ & $2$&$.5$ \\

$3$ & $b=0$ & $1$&$.145$ & $0$& & $0$&$.144$ & $0$&$.981$ & $-0$&$.066\,16$ &
$5$&$.4$ \\

$3$ & $b=0$, $\gamma=1$ & $1$&$.140$ & $0$& & $0$&$.150$ & $1$& &
$-0$&$.066\,24$ & $44$&$.7$ \\

$3$ & $a=b$ & $0$&$.774$ & $0$&$.774$ & $-0$&$.466$ & $0$&$.002$ &
$-0$&$.523\,64$ & $2\,840$& \\

$3$ & $a=\gamma=1$, $b=0$ & $1$& & $0$& & $1$& & $-0$&$.125$ & $-0$&$.066\,57$
& $26\,300$& \\

$3$ & Eq.\ (\ref{eqn:extrap_formula_ceperley_1978}) &
\multicolumn{2}{c}{$\cdots$} & \multicolumn{2}{c}{$\cdots$} & $0$&$.331$ &
\multicolumn{2}{c}{$\cdots$} & $-0$&$.065\,55$ & $40\,800$& \\

$8$ & None & $1$&$.218$ & $-0$&$.010$ & $0$&$.056$ & $1$&$.062$ &
$-0$&$.061\,21$ & $21$&$.7$ \\

$8$ & $b=0$ & $1$&$.204$ & $0$& & $0$&$.056$ & $1$&$.073$ & $-0$&$.061\,21$ &
$21$&$.8$ \\

$8$ & $b=0$, $\gamma=1$ & $1$&$.246$ & $0$& & $0$&$.048$ & $1$& &
$-0$&$.061\,12$ & $377$& \\

$8$ & $a=\gamma=1$, $b=0$ & $1$& & $0$& & $1$& & $-0$&$.004$ & $-0$&$.061\,19$
& $10\,600$& \\

$8$ & Eq.\ (\ref{eqn:extrap_formula_ceperley_1978}) &
\multicolumn{2}{c}{$\cdots$} & \multicolumn{2}{c}{$\cdots$} & $0$&$.092$ &
\multicolumn{2}{c}{$\cdots$} & $-0$&$.060\,92$ & $22\,600$& \\

\hline \hline
\end{tabular}
\caption{Results of fitting Eqs.\ (\ref{eqn:heg_fs_fit_formula}) and
(\ref{eqn:extrap_formula_ceperley_1978}) to non-twist-averaged DMC
energy data for a paramagnetic Fermi fluid.  Seven different system
sizes ($N=18$, 54, 118, 226, 338, 458, and 566) were used for each
density, and the statistical error bars in the total energy were
around $0.00001$--$0.00005$ a.u.\ per electron.  Time steps of
$0.0015$, $0.033$, and $0.1$ a.u.\ were used in the simulations at
$r_s=1$, $3$, and $8$ a.u., respectively.  The target population was
3200 configurations in each case.  The wave-function was of
Slater-Jastrow form, i.e., backflow was not used.  The constraint
$a=b$ leads to an enormous $\chi^2$ value at $r_s=8$ a.u.  The
infinite-system DMC energies from the twist-averaged DMC calculations
at $r_s=1$ and 3 a.u.\ are $0.5880(6)$ and $-0.06623(3)$ a.u.\ per
electron, respectively. Twist-averaged calculations have not been
performed at $r_s=8$ a.u.  \label{table_DMC_fs_fits}}
\end{center}
\end{table}

\section{Size-dependence of biases in DMC energies \label{sec:biases}}

\begin{figure}
\begin{center}
\includegraphics[scale=0.3,clip]{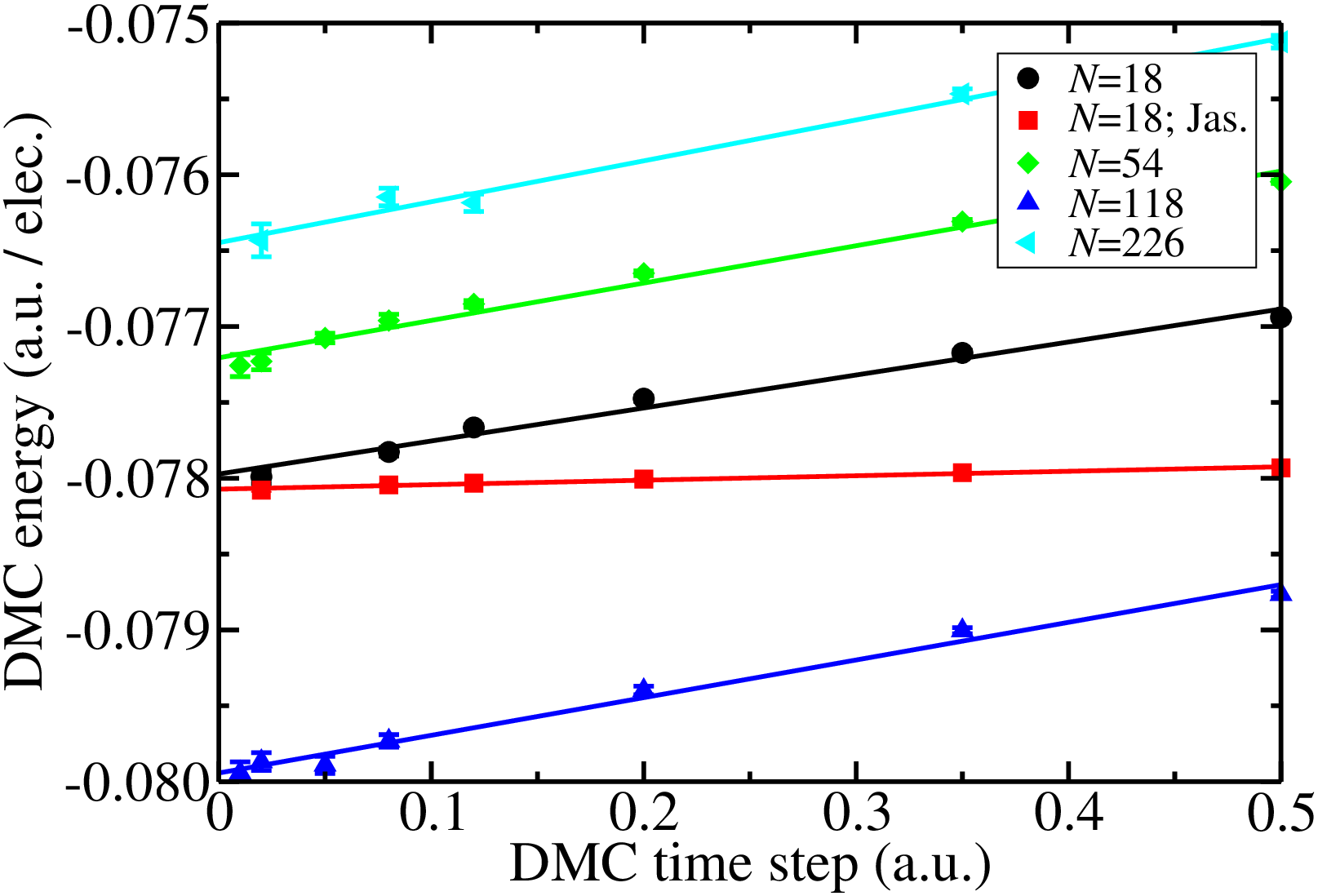}
\end{center}
\caption{(Color online) DMC energy per electron against time step for a
paramagnetic 3D HEG of density parameter $r_s=4$ a.u.\ at various system sizes
$N$.  A Slater wave function was used, except for the one curve labeled
``Jas,'' in which a Slater-Jastrow wave function was used.  Twist averaging
was not applied.
\label{fig:HEG_rs4_para_dt_bias}}
\end{figure}

Figure \ref{fig:HEG_rs4_para_dt_bias} shows that the time-step bias in the DMC
energy per particle has nearly the same form over the range of system sizes
typically encountered in DMC simulations.  A time step judged to be accurate
in a small system should therefore continue to be accurate in a larger system.
To exaggerate the bias, most of the results shown in Fig.\
\ref{fig:HEG_rs4_para_dt_bias} were obtained using a simple Slater trial
function with no Jastrow factor; the bias is greatly reduced if a more
accurate trial wave function is used, as can also be seen in Fig.\
\ref{fig:HEG_rs4_para_dt_bias}.

\begin{figure}
\begin{center}
\includegraphics[scale=0.3,clip]{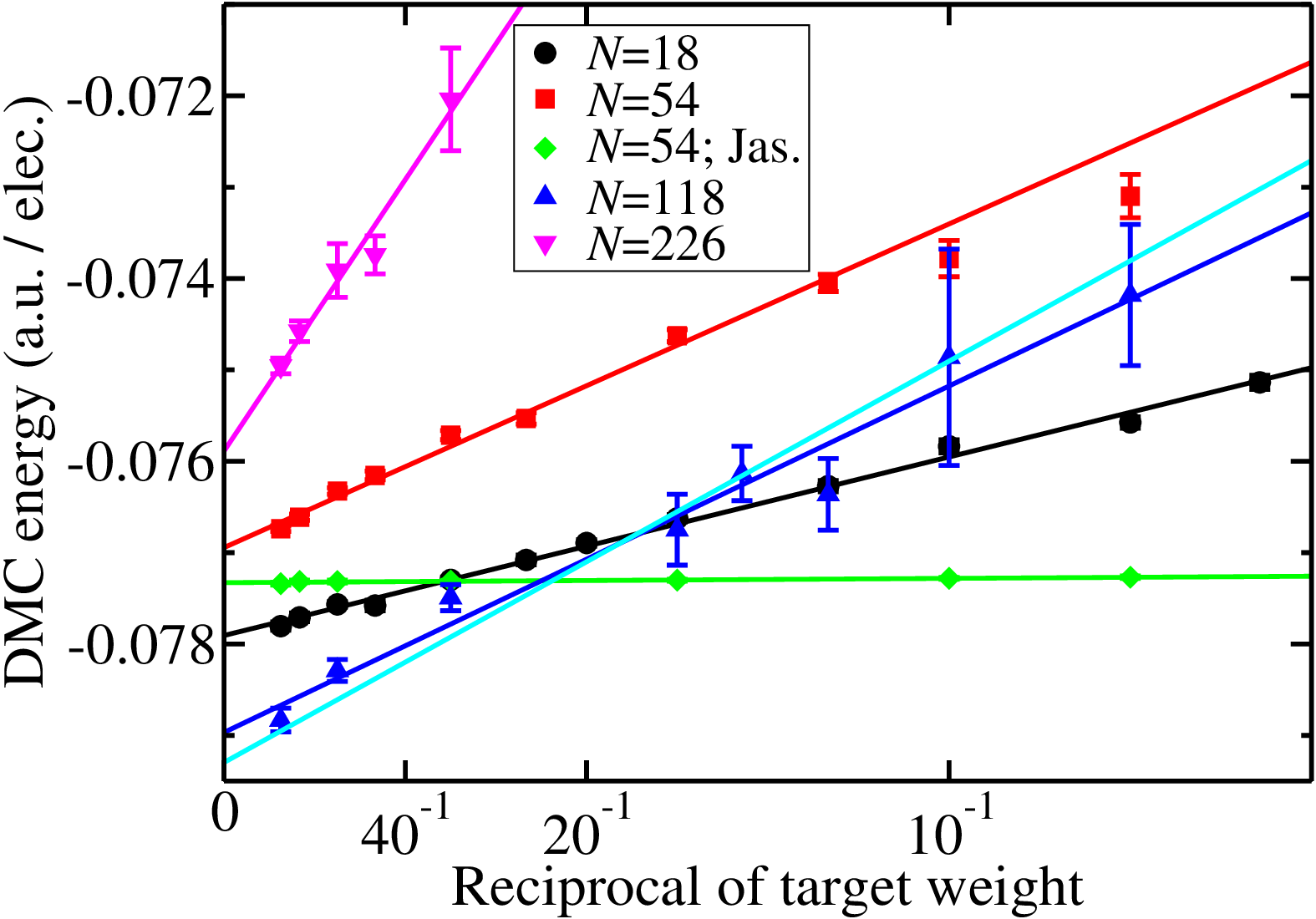} \\[1ex]
\includegraphics[scale=0.3,clip]{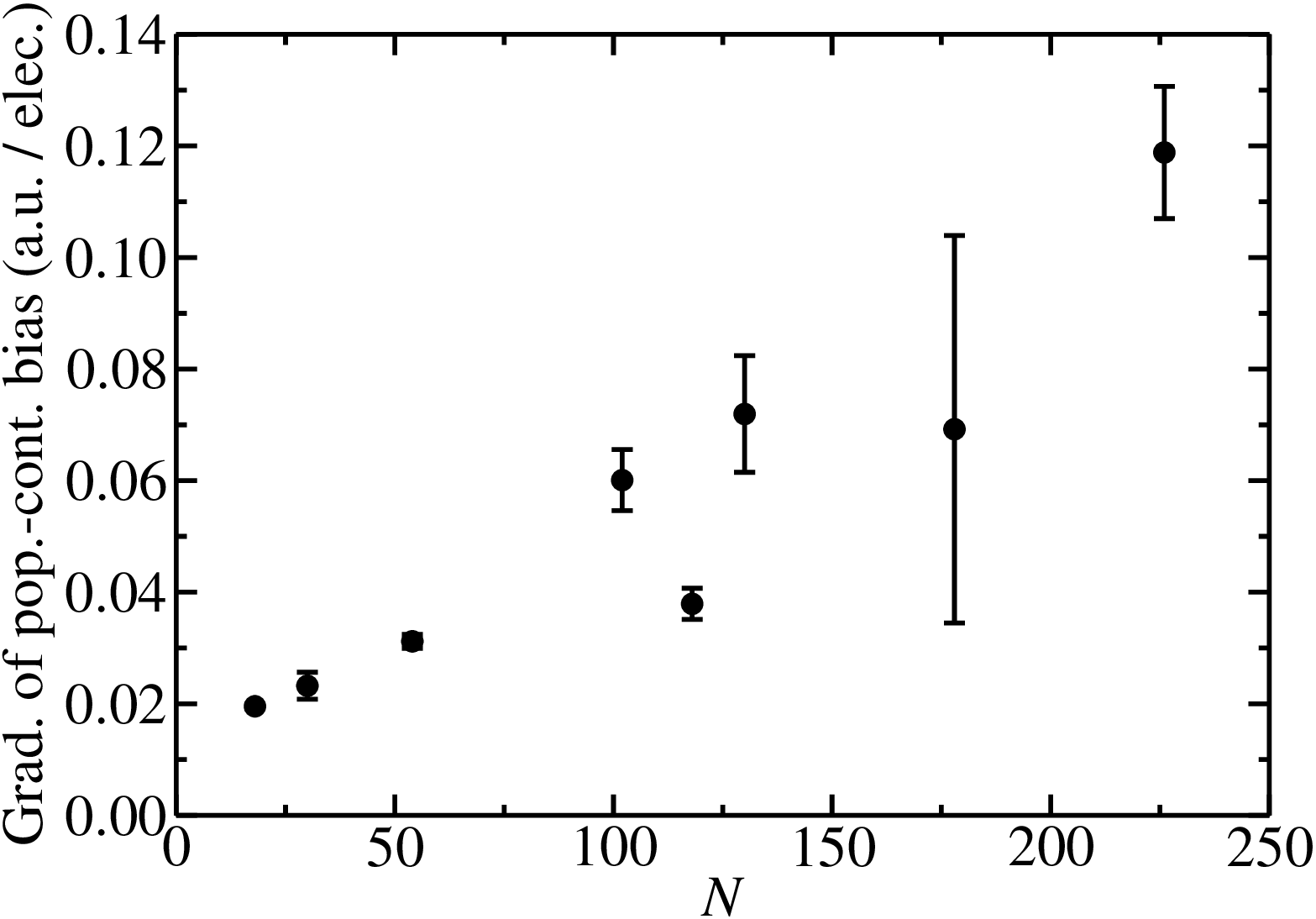}
\end{center}
\caption{(Color online) Upper panel: DMC energy per electron against
reciprocal of target population for a paramagnetic 3D HEG of density parameter
$r_s=4$ a.u.\ at various system sizes $N$.  Lower panel: gradient of the
population-control bias (derivative of the DMC energy per electron with
respect to the reciprocal of the target population) against system size.  The
DMC time step was 0.03 a.u., and a Slater wave function was used, except for
the one curve in the top panel labeled ``Jas,'' in which a Slater-Jastrow wave
function was used.  Twist averaging was not applied.
\label{fig:HEG_rs4_para_Nc_bias}}
\end{figure}

For any given system, the DMC population-control bias should fall off roughly
as $N_C^{-1}$, where $N_C$ is the target population,\cite{umrigar_1993} so we
have plotted the DMC energy against $N_C^{-1}$ in Fig.\
\ref{fig:HEG_rs4_para_Nc_bias}. Unlike time-step bias, population-control bias
grows with system size. However, the increase in the bias with system size is
slow.  Population-control bias is caused by the correlation of fluctuations in
the local energy and the DMC branching factor.\cite{umrigar_1993} Fluctuations
in the local energy increase as $N^{1/2}$.  If the exponential branching
factors can be approximated by the first two terms in the Taylor expansion of
the exponential then fluctuations in the branching factor increase as
$N^{1/2}$.  So the population-control bias in the energy per particle is
roughly independent of system size.  However, the fluctuations in the
exponential branching factor grow more rapidly than $N^{1/2}$ in large
systems, causing the bias to increase.  Improving the accuracy of the trial
wave function reduces population-control bias, as can be seen in the upper
panel of Fig.\ \ref{fig:HEG_rs4_para_Nc_bias}.

\section{Conclusions \label{sec:conclusions}}

We have carried out a detailed study of finite-size effects in QMC
calculations and have described a number of approaches for reducing or
correcting them.  Twist averaging greatly reduces the magnitude of
single-particle finite-size errors, although residual single-particle errors
due to the wrong shape of the CE twist-averaged Fermi surface are still
significant in studies of the HEG\@. One can calculate these errors within HF
theory, and hence correct for them.

Finite-size effects in the XC energy should be eliminated, either by adding a
correction to the Ewald energy or by using the MPC interaction to calculate
the final energies (although the Ewald interaction should be used to generate
the configuration distribution, since the MPC interaction distorts the XC hole
in finite systems).  Finite-size corrections must also be applied to the KE\@.
For HEGs, where analytic expressions for the low-$k$ behavior of the two-body
Jastrow factor are available, we have found that it is important to include
both the leading- and next-to-leading-order KE corrections at intermediate and
high densities.  The resulting QMC energy data are almost independent of
particle number at typical system sizes.  For real systems we recommend
fitting the QMC-optimized Jastrow factor to Eq.\ (\ref{eqn:approx_urpa}) at
small $k$, then using Eq.\ (\ref{eqn:gen:_KE_corr}) to compute the correction
to the KE\@.

Within HF theory the long-ranged nature of the exchange hole leads to
additional errors in the exchange energy.  These errors are absent in QMC
calculations.  They can also be viewed as arising from the nonanalytic
behavior of the HF structure factor at ${\bf k}={\bf 0}$.  We have constructed
an accurate correction for these errors in HF theory.

For 2D systems the leading-order finite-size errors (using both the Ewald and
MPC interactions) are caused by the slow convergence of the XC hole and the
neglect of long-ranged correlations in the KE\@.  The errors in the energy per
particle scale as $O(N^{-5/4})$, suggesting that this form should be assumed
in the extrapolation to infinite system size.

If the single-particle finite-size error is to be removed by extrapolation
rather than twist averaging then the HF exchange energy should not be included
in the extrapolation; just the KE\@.  Furthermore, an estimate of the
effective mass should be included in the extrapolation.

Tests at realistic system sizes show that time-step bias in DMC results does
not get significantly worse as the system size is increased.  Population
control bias does get worse, but only slowly.

\section{Acknowledgments}

Financial support has been provided by Jesus College, Cambridge, and the
Engineering and Physical Sciences Research Council (EPSRC), UK\@.  Computing
resources have been provided by the Cambridge High Performance Computing
Service, the Imperial College High Performance Computing Service, and the UK
National HPCx service. We thank D.\ M.\ Ceperley and M.\ Holzmann for helpful
conversations.

\appendix

\section{Finite-size errors in HF theory \label{sec:hf_fs_errors}}

For the 3D HEG, the HF exchange hole is\cite{giuliani}
\begin{eqnarray}
\rho_{\rm x}(r) & = & - \frac{1}{N\Omega} \sum_{\sigma} N_{\sigma}^2  \left| 3
\frac{\sin(k_{{\rm F} \sigma} r)-k_{{\rm F} \sigma} r  \cos(k_{{\rm F} \sigma}
r)}{(k_{{\rm F} \sigma} r)^3} \right|^2 \nonumber \\ & \simeq & - \frac{9}{2 N
\Omega r^4} \left( \frac{\Omega}{6 \pi^2} \right)^{4/3} \sum_{\sigma}
N_{\sigma}^{2/3} ,
\end{eqnarray}
where in the last line we have retained only the dominant nonoscillatory term
at large separation, $k_{{\rm F} \sigma}=\left( 6 \pi^2 N_{\sigma} / \Omega
\right)^{1/3}$ is the Fermi wave vector for particles of spin $\sigma$, and
$N_{\sigma}$ is the number of particles of spin $\sigma$.  The hole has a
slowly decaying tail that falls off as $1/r^4$, so there is a missing
contribution to the exchange energy in a finite simulation cell. The
interaction of each electron with its exchange hole should be $1/r$ (as
enforced inside the simulation cell when the MPC interaction is used). So the
missing contribution to the HF interaction energy is approximately
\begin{eqnarray}
\Delta V_{\rm HF}^{\rm (1)} & = & \frac{N}{2} \int_{R_\Omega}^\infty \frac{4
  \pi r^2 \rho_{\rm x}(r)}{r} \, dr \nonumber \\  & = & - \frac{1}{2 \pi r_s}
  \left( \frac{3}{4 \pi N} \right)^{1/3}  \sum_{\sigma} N_{\sigma}^{2/3},
\end{eqnarray}
where $R_\Omega$ is the radius of a sphere of volume $\Omega$.  This gives a
finite-size error in the HF energy per particle that falls off slowly as
$N^{-2/3}$.  This error will also be present in the Ewald energy.  In addition
to this missing contribution, there are errors arising from the fact that the
part of the exchange hole that would lie outside the simulation cell if the
system were infinite is distorted by being compressed back into the simulation
cell to satisfy the sum rule.  The charge of the missing tail is approximately
\begin{equation} Q=\int_{R_\Omega}^\infty 4\pi r^2 \rho_{\rm x}(r) \, dr =
  \frac{-3}{N\pi}\left( \frac{\Omega}{6\pi^2}\right)^{1/3} \sum_{\sigma}
  N_{\sigma}^{2/3} \frac{1}{R_\Omega} = \frac{-3}{N \pi} \left(\frac{2}{9 \pi}
  \right)^{1/3} \sum_{\sigma} N_{\sigma}^{2/3}.
\end{equation}
If we assume that this missing charge is uniformly distributed inside a sphere
of radius $R_\Omega$, we must subtract its unwanted contribution to the
exchange energy, giving another correction
\begin{equation} \Delta V_{\rm HF}^{\rm (2)}=-\frac{3NQ}{4R_\Omega} =
  \frac{3}{2\pi r_s} \left( \frac{3}{4 \pi N} \right)^{1/3}  \sum_{\sigma}
  N_{\sigma}^{2/3}.
\end{equation}
(Other approximations, such as assuming $Q$ to increase linearly within
$R_\Omega$ may be more accurate.)  The total correction to the exchange energy
(either Ewald or MPC) obtained within this real-space procedure is
\begin{equation}
\Delta V_{\rm HF} = \Delta V_{\rm HF}^{\rm (1)} + \Delta V_{\rm HF}^{\rm (2)}
  = \frac{1}{\pi r_s} \left( \frac{3}{4 \pi N} \right)^{1/3} \sum_{\sigma}
  N_{\sigma}^{2/3}.
  \label{eqn:hf_corr_X_A}
\end{equation}
The result of applying this correction to the HF Ewald exchange energy is
shown in Fig.\ \ref{fig:HF_X_energy}, along with the result of applying the
correction of Eq.\ (\ref{eqn:hf_corr_X_B}).  Both work well, although the
correction of Eq.\ (\ref{eqn:hf_corr_X_B}) is better.

\section{Equivalence of the MPC and XC correction \label{app:corr_mpc_equiv}}

Consider a system of cubic symmetry.  The difference between the MPC and Ewald
XC energies is:
\begin{eqnarray}
\langle \hat{V}_{\rm MPC} \rangle - \langle \hat{V}_{\rm Ew} \rangle & = &
\frac{N}{2}\int_{\Omega} \rho_{\rm xc}({\bf r}) \left( \frac{1}{r} - [v_E({\bf
r}) - v_M] \right) \, d{\bf r} \nonumber \\ & = & \frac{N}{2}\int_{\Omega}
\rho_{\rm xc}({\bf r}) \left( \frac{2\pi}{3\Omega} r^2 + \ldots \right) \,
d{\bf r} ,
\end{eqnarray}
were we have used the expansion of the Ewald interaction from Eq.\
(\ref{eqn:taylor_ewald}). Assuming that $\rho_{\rm xc}({\bf r})$ is well
localized within the simulation cell, we can replace $\rho_{\rm xc}({\bf r})$
by $S_{\rm loc}({\bf r}) - \delta({\bf r})$ and extend the range of
integration to infinity to obtain
\begin{eqnarray}
\langle \hat{V}_{\rm MPC} \rangle - \langle \hat{V}_{\rm Ew} \rangle & = &
\frac{N\pi}{3\Omega} \int_{r < \infty} r^2 S_{\rm loc}({\bf r}) \, d{\bf r}
\nonumber \\ & = & - \frac{N\pi}{3\Omega} \left.  \nabla_{\bf k}^2 S_{\rm
loc}({\bf k}) \right |_{{\bf k} = {\bf 0}}.
\end{eqnarray}
Since $S({\bf k}) = \eta k^2 + O(k^4)$ in a cubic system, this reproduces Eq.\
(\ref{eq:finsize1}):
\begin{equation}
\langle \hat{V}_{\rm MPC} \rangle - \langle \hat{V}_{\rm Ew} \rangle =
\frac{2\pi N \eta}{\Omega} + \ldots .
\end{equation}
The use of the first-order $\Delta V_{\rm Ew}$ correction may therefore be
regarded as a first-order approximation to the MPC, in which the leading term
in the small-$r$ expansion of the difference between $1/r$ and $v_E({\bf r}) -
v_M$ is taken into account but higher order terms are neglected.


\begin{thebibliography}{99}
\bibitem{footnote:xc_energy} Note that the definition of the XC energy given
  here (the interaction of each electron with its XC hole) differs from the
  definition of the XC energy given in some other contexts (the difference of
  the energy within Hartree theory and the ground-state energy).
\bibitem{footnote:shell_spherical} Note that, in general, the average of the
  SF over a shell of symmetry-equivalent ${\bf G}_s$ vectors is not the same
  as the spherical average of the SF evaluated at radius $|{\bf G}_s|$.
\bibitem{footnote:k3_term} If $S(k)$ includes an $O(k^3)$ term then the error
  in the XC-corrected Ewald energy falls off as $N^{-1/3}$.
\bibitem{footnote_effmass} The correction for the residual single-particle
  finite-size error from HF theory should be divided by the effective electron
  mass.  This can be estimated as the ratio of the HF KE to the QMC KE\@.
  However, since the residual single-particle error is already a small
  correction, it makes little difference to the final results if one assumes
  an effective mass of 1 a.u.
\end{thebibliography}
\end{document}